\documentclass[twocolumn,prb,showpacs,aps,superscriptaddress]{revtex4}

\usepackage{amsmath}
\usepackage{amssymb,amsthm}
\usepackage{mathptmx}
\usepackage{graphicx}
\usepackage{color}

\DeclareMathAlphabet{\mathcal}{OMS}{cmsy}{m}{n}

\newcommand{\cds}[1]{\ensuremath{c^\dagger_{#1}}}
\newcommand{\ccs}[1]{\ensuremath{c_{#1}}}
\newcommand{\ads}[1]{\ensuremath{a^\dagger_{#1}}}
\newcommand{\aas}[1]{\ensuremath{a_{#1}}}
\newcommand{\cd}[1]{\ensuremath{\hat{c}^\dagger_{#1}}}
\newcommand{\cc}[1]{\ensuremath{\hat{c}_{#1}}}

\newcommand{\lwf}[1]{\ensuremath{\langle #1 \vert}}
\newcommand{\rwf}[1]{\ensuremath{\vert #1 \rangle}}
\newcommand{\est}[1]{\ensuremath{\langle {#1} \rangle}}
\newcommand{\dt}[0]{\ensuremath{\mathrm{d}t}}
\renewcommand{\exp}[1]{\ensuremath{\mathrm{exp}\left({#1}\right)}}
\newcommand{\trace}[1]{\ensuremath{\mathrm{tr}\left({#1}\right)}}
\newcommand{\torder}[0]{\ensuremath{\mathcal{T}}}
\newcommand{\tcorder}[0]{\ensuremath{\mathcal{T}_C}}

\begin{document}

\title{Lehmann representation of the nonequilibrium self-energy}

\author{Christian Gramsch}
\author{Michael Potthoff}
\affiliation{I. Institute for Theoretical Physics, University of Hamburg,
  Jungiusstra\ss e 9, 20355 Hamburg, Germany}
\affiliation{The Hamburg Centre for Ultrafast Imaging, Luruper Chaussee 149,
  22761 Hamburg, Germany}

\begin{abstract}
  It is shown that the nonequilibrium self-energy of an interacting
  lattice-fermion model has a unique Lehmann representation.  Based on the
  construction of a suitable non-interacting effective medium, we provide an
  explicit and numerically practicable scheme to construct the Lehmann
  representation for the self-energy, given the Lehmann representation of the
  single-particle nonequilibrium Green's function.  This is of particular
  importance for an efficient numerical solution of Dyson's equation in the
  context of approximations where the self-energy is obtained from a reference
  system with a small Hilbert space. As compared to conventional techniques
  to solve Dyson's equation on the Keldysh contour, the effective-medium
  approach allows to reach a maximum propagation time which can be several
  orders of magnitude longer.  This is demonstrated explicitly by choosing the
  nonequilibrium cluster-perturbation theory as a simple approach to study the
  long-time dynamics of an inhomogeneous initial state after a quantum quench
  in the Hubbard model on a $10 \times 10$ square lattice.  We demonstrate that
  the violation of conservation laws is moderate for weak Hubbard interaction
  and that the cluster approach is able to describe prethermalization physics. 
\end{abstract}



\maketitle

\section{Introduction}
The study of physical phenomena that arise in strongly correlated systems far
from equilibrium has become a field of highly active research recently.
\cite{pol:11, aoki:14} For the theoretical description of such systems,
Green's-function-based approaches starting from the Keldysh formalism
\cite{kel:64} have proven to be very useful. A number of different
approximation schemes rely on this concept. \cite{thy:07, sch:02, fre:06,
  bal:05.11, jung:11, knap:11, hof:13, jou:15} Central to these approaches is
the self-energy which is related to the one-particle Green's function through
Dyson's equation. However, while the numerical solution of Dyson's equation is
rather straightforward in the equilibrium case, the computational effort is
considerably increased for systems out of equilibrium since operations with
matrices depending on two independent contour time variables typically scale
cubically in the number of time steps.  Apart from other challenges
characteristic for the respective approach, already this scaling poses a severe
limit on the maximal reachable propagation time in a numerical calculation.
Applying additional concepts or approximations, such as the generalized
Kadanoff-Baym ansatz\cite{lip:86, bo:14} or exploiting a rapid decay of the
memory,\cite{thor:08} are necessary to overcome this limitation.

It was proposed recently\cite{eck:01.14} that it can be advantageous to avoid
the direct inversion of Dyson's equation by applying a mapping onto a Markovian
propagation scheme. To this end it is necessary to {\em assume} the existence
of a certain functional form for the nonequilibrium self-energy, namely the
existence of a Lehmann representation. 

In the present paper we explicitly construct this Lehmann representation. With
this at hand, we pick up the proposed idea to solve Dyson's equation by means
of a Markovian propagation and exploit the fact that the Lehmann representation
of the exact self-energy of a small reference system has a finite number of
terms only. This allows us to solve Dyson's equation with an effort that scales
linearly in the maximum propagation time $t_\mathrm{max}$.

For equilibrium Green's functions, the Lehmann representation is a well
established concept.\cite{wal:81} It uncovers the analytical properties of the
Green's function and can be used to show that the related spectral function is
positive definite. It is further essential for the evaluation of diagrams
through contour integrations in the complex frequency plane, for the derivation
of sum rules, etc.  The generalization of the Lehmann representation to
nonequilibrium Green's function is straightforward.\cite{cg:13} Applications
include nonequilibrium dynamical mean-field theory (DMFT) where it allows for a
Hamiltonian-based formulation of the impurity problem.\cite{cg:13}

The explicit construction of a Lehmann representation for the {\em
  self-energy}, on the other hand, turns out to be more tedious, already for
the equilibrium case: In a recent work such a construction was worked out
\cite{ste:14} from a diagrammatic perspective and used to cure the problem of
possibly negative spectral functions arising from a summation of a subclass of
diagrams. 

Here, we address the nonequilibrium self-energy of a general, interacting
lattice-fermion model: (i) We rigorously show the existence of the Lehmann
representation by presenting an explicit construction scheme that is based on
the Lehmann representation of the nonequilibrium Green's function.  (ii) Using
a simple example, namely the cluster-perturbation theory \cite{gros:93, sen:00,
  sen:02, bal:05.11, knap:11, jur:13} (CPT), we furthermore demonstrate that
the Lehmann representation of the self-energy can in fact be implemented
numerically and used to study the time evolution of a locally perturbed Hubbard
model on a large square lattice ($10 \times 10$ sites).  Propagation times of
several orders of magnitude in units of the inverse hopping amplitude can be
reached with modest computational resources.  (iii) While the CPT approximation
for the self-energy is rather crude and shown to violate a number of
conservation laws, it is possible with this approximation to study the
weak-coupling limit of the Hubbard model in a reasonable way. In particular we
demonstrate that prethermalization physics is already captured on this level.

The paper is organized as follows: In Section \ref{sec:repg} we briefly discuss
the generalization of the Lehmann representation to nonequilibrium Green's
functions.  The main idea of Ref.\ \onlinecite{eck:01.14} about the Markovian
propagation scheme is recalled in Sec.\ \ref{sec:prevres}. The explicit
construction scheme for the nonequilibrium self-energy is outlined in Section
\ref{sec:expl_const}.  Section \ref{sec:appl_cpt} is devoted to the application
of our formalism to the cluster-perturbation theory. Sec.\ \ref{sec:numerics}
presents numerical results for the time evolution of a local perturbation in
the fermionic Hubbard model.  We conclude the paper with a summary and an
outlook in Sec.\ \ref{sec:summary}.

\section{Lehmann representation of the one-particle Green's function}
\label{sec:repg}
We consider an arbitrary, fermionic model Hamiltonian
\begin{equation}
  \label{eq:ham}
  H(t) =   \sum_{ij} (T_{ij}(t)-\delta_{ij}\mu) \cds{i}\ccs{j} 
         + \frac{1}{2}\sum_{iji'j'}U_{ii'jj'}(t)\cds{i}\cds{i'}\ccs{j'}\ccs{j},
\end{equation}
where the indices $i,j$ run over the possible one-particle orbitals (lattice
sites, local orbitals, spin projection, ...). Fermions in such states are
created (annihilated) by the operators $\cds{i}$ ($\ccs{i}$).  At time $t=0$,
the system with Hamiltonian $H(0) = H_\mathrm{ini}$ is assumed to be in thermal
equilibrium with inverse temperature $\beta$ and chemical potential $\mu$.
Non-equilibrium real-time dynamics for $t>0$ is initiated by the time
dependence of the one-particle or the interaction parameters. 
This covers challenging experimental setups such as 
time-resolved photoemission spectroscopy\cite{perf:06} or experiments with
ultracold gases in optical lattices.\cite{bloch:08}

\begin{figure}[t]
  \includegraphics[width=0.4\textwidth]{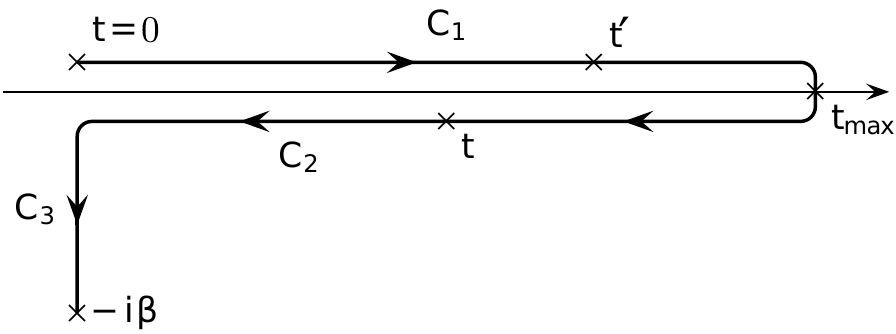}
  \caption{Keldysh-Matsubara contour $C$. $C_1$ denotes the upper branch, $C_2$
    the lower branch and $C_3$ the Matsubara branch. In the shown example $t$
    is later than $t'$ in sense of the contour, denoted as $t>_C t'$ in the
    text.}
  \label{fig:contour}
\end{figure}

The one-particle Green's function is given by
\begin{align}
	\label{eq:gf}
  G_{ij}(t,t') &=   -i\est{\tcorder\,\cc{i}(t)\cd{j}(t')}_{H}
                  \nonumber \\
               &\equiv\frac{-i}{Z}\trace{\exp{-\beta
                   H_\mathrm{ini}}\left[\tcorder\,\cc{i}(t)\cd{j}(t')\right]},
\end{align}
where ``$\text{tr}(\dots)$'' traces over the Fock space, i.e., we take averages
using the grand-canonical ensemble. \makebox{$Z=\trace{\exp{-\beta 
H_\mathrm{ini}}}$} defines the grand-canonical partition function and
$\tcorder$ the time-ordering operator on the L-shaped Keldysh-Matsubara contour
$C$ (\makebox{see Fig.\ \ref{fig:contour}}).  The time variables $t$ and $t'$
are understood as contour times that can lie on the upper, lower or Matsubara
branch of $C$.  We further introduce the convention that operators with a hat
carry a time dependence according to the Heisenberg picture, i.e.,\ $\cc{i}(t)
= U^\dagger(t,0)\ccs{i}U(t, 0)$, where $U(t,t') = \torder \exp{-i\int^t_{t'}
  H(t_1)\dt_1}$ is the system's time-evolution operator and $\torder$ the
time-ordering operator. An in-depth introduction to the Keldysh formalism
\cite{kel:64} can be found in Refs.\ \onlinecite{ram:07}, \onlinecite{lee:06}.

As has been shown in Ref.\ \onlinecite{cg:13}, the one-particle Green's
function can be cast into the form
\begin{equation}
	\label{eq:representable}
	G_{ij}(t,t')=\sum_{\alpha}Q_{i\alpha}(t)
                  g(\epsilon_\alpha; t,t')Q_{j\alpha}^*(t'),
\end{equation}
which we will call its Lehmann representation in the following.
$g(\epsilon;t,t')$ is the non-interacting Green's function of an isolated
one-particle mode ($h_\mathrm{mode}=\epsilon \cds{}\ccs{}$) with excitation
energy $\epsilon$:
\begin{equation}
  \label{eq:isolated}
  g(\epsilon;t,t')=i[f(\epsilon)-\Theta_C(t,t')] e^{-i\epsilon(t-t')}.
\end{equation}
Here, $f(\epsilon)=(e^{\beta \epsilon}+1)^{-1}$ denotes the Fermi-function
while $\Theta_C(t,t')$ refers to the contour variant of the Heaviside step
function ($\Theta_C(t,t')=1$ for $t\ge_Ct'$, $\Theta_C(t,t')=0$ otherwise).
$Q(t)$ is defined to be equal on the upper and lower branch of the contour and
furthermore constant on the Matsubara branch with $Q(-i\tau)=Q(0)$ and 
$\tau\in[0,\beta]$.  If the eigenstates $\rwf{m}$ of the initial Hamiltonian
(i.e.,  $H_\mathrm{ini}\rwf{m} = E_m \rwf{m}$) are used as a basis for tracing
over the Fock space in Eq.\ \eqref{eq:gf}, one has
\begin{align}
  \label{eq:qmat}
	Q_{i\alpha}(t)  &= Q_{i(m,n)}(t)
                   = z_{(m,n)}
                      \lwf{m}\cc{i}(t)\rwf{n}e^{i\epsilon_{(m,n)}t},
\end{align}
where $z_{(m,n)} = \sqrt{(e^{-\beta E_m}+e^{-\beta E_n})/Z}$ and where the
superindex $\alpha=(m,n)$ labels the possible one-particle excitations with
corresponding excitation energies $\epsilon_\alpha=\epsilon_{(m,n)} = E_n -
E_m$.  Note that this definition of $Q(t)$ indeed satisfies $Q(-i\tau)=Q(0)$.
We emphasize that $Q$ as a matrix is not quadratic. Our expression can be seen
as a direct generalization of the time-independent $Q$-matrix discussed in
Ref.\ \onlinecite{aich:06}.  We further note that the rows of the $Q$-matrix
fulfill the orthonormality condition 
\begin{equation}
  \label{eq:orthonormal}
  [Q(t) Q^\dagger(t)]_{ij}=\sum_\alpha Q_{i\alpha}(t)Q_{j\alpha}^*(t)
      = \est{\left\{\cc{i}(t),\cd{j}(t)\right\}}_{H}=\delta_{ij},
\end{equation}
where $\left\{A,B\right\} = AB + BA$ denotes the anticommutator. 

\section{Lehmann representation of the self-energy}
\label{sec:lsig}
\subsection{Motivation}
\label{sec:prevres}
In several Green's-function-based methods, an approximate
self-energy $\Sigma'$ is obtained from a small reference system using exact
diagonalization. The desired one-particle Green's function $G$ of a much larger
system is then obtained through Dyson's equation
\begin{align}
  \label{eq:dyson_volterra}
  G_{ij}(t,t')&=  [G_0]_{ij}(t,t') \\
  &\,\,\,+ \int_C\int_C \dt_1\dt_2 
  \sum_{k_1k_2}[G_0]_{ik_1}(t,t_1)\Sigma'_{k_1k_2}(t_1,t_2)
  G_{k_2j}(t_1,t'),\nonumber
\end{align}
where $G_0$ denotes the non-interacting Green's function (i.e., $U=0$) of the
model given by Eq.\ \eqref{eq:ham}. Typical examples include dynamical
mean-field theory (DMFT), \cite{voll:01.89, kot:03.92, sch:02, fre:06} where
$\Sigma'$ is obtained from a single-impurity Anderson model,\cite{cg:13} or
cluster-perturbation \cite{gros:93, sen:00, sen:02, bal:05.11, knap:11, jur:13}
and self-energy functional theory, \cite{po:03,hof:13,hof:15} where $\Sigma'$ stems
from a small reference system. To solve Eq.\ \eqref{eq:dyson_volterra}
numerically, a discretization of the continuous time-contour $C$ is necessary.
The number of time steps required to reach a given maximal time is dependent on
the lowest relevant timescale that is set by a given Hamiltonian. Based on this
discretization, the effort required to solve Eq.\ \eqref{eq:dyson_volterra} for
$G$ scales cubically in the number of time steps and also the system size.
Despite this challenge also the memory consumption, which scales quadratically
in these quantities, poses a problem.  Progress was made recently
\cite{eck:01.14} by introducing a mapping of Eq.\ \eqref{eq:dyson_volterra}
onto a Markovian propagation-scheme. 

The idea proposed by the authors of Ref.\ \onlinecite{eck:01.14} relies on the
\emph{assumption} that the self-energy can be written in the following form: 
\begin{equation}
  \label{sig:leh}
  \Sigma'_{ij}(t,t')=  \delta_C(t,t') \Sigma'{}^\mathrm{HF}_{ij}(t)
  + \sum_{s} h_{is}(t) g(h_{ss};t,t') h^*_{js}(t).
\end{equation}
Here, ${\Sigma'}_{ij}^\mathrm{HF}(t)$ denotes the time-local Hartree-Fock term.
This decomposition is very similar to the expression Eq.\
\eqref{eq:representable} for the Green's function.  We will refer to this as
the Lehmann representation of the self-energy. The immediate and important
advantage of the Lehmann representation is that the self-energy can be
interpreted as a hybridization function.\cite{cg:13, eck:01.14} This
property allows to write down an effective non-interacting model with
Hamiltonian 
\begin{align}
  \label{eq:heff_eck}
  H_\mathrm{eff}(t) =   \sum_{ij} &(T_{ij}(t) 
                    + \Sigma'{}^\mathrm{HF}_{ij}(t))\cds{i}\ccs{j}\\
                    +&\sum_{is} (h_{is}(t) \cds{i}\aas{s} + \mathrm{h.c.})
                    + \sum_s h_{ss} \ads{s}\aas{s}\nonumber.
\end{align}
The $s$-degrees of freedom represent ``virtual'' orbitals in addition to the
physical degrees of freedom labeled by $i$. They form an ``effective
medium'' with on-site energies $h_{ss}$ and hybridization strengths $h_{is}(t)$
such that the {\em interacting} Green's function of the original model is the
same as the Green's function of the effective non-interacting model on the
physical orbitals:
\begin{equation}
  G_{ij}(t,t')=-i\est{\tcorder\,\cc{i}(t)\cd{j}(t')}_{H_\mathrm{eff}}.
\end{equation}
With this simple construction, the inversion of the Dyson equation can be
avoided in favor of a Markovian time propagation within a non-interacting
model.

As a successful benchmark, an interaction quench in an inhomogeneous Hubbard
model was treated with nonequilibrium DMFT in Ref. \onlinecite{eck:01.14} using
self-consistent second-order perturbation theory as impurity solver.  On the
theoretical side, however, it remained an open question if the existence of a
Lehmann representation must be postulated or if this is a general property of
the nonequilibrium self-energy.  

In the following we explicitly derive Eq.\ \eqref{sig:leh} for the exact
self-energy corresponding to the general, interacting Hamiltonian defined in
Eq.\ \eqref{eq:ham}, i.e., we show that the exact self-energy can always be
written in the form of a Lehmann representation.  The proposed construction
scheme is not only useful as an analytical tool but also well suited for
numerical applications where an approximate self-energy is obtained from a
small reference system using exact diagonalization. In this case the number of
virtual orbitals is constant and the effort for solving Eq.\
\eqref{eq:dyson_volterra} scales linearly in $t_\mathrm{max}$. This is a great
advantage if one is interested in long-time dynamics.

\subsection{Explicit construction}
\label{sec:expl_const}
\begin{figure}[t!]
  \includegraphics[width=0.45\textwidth]{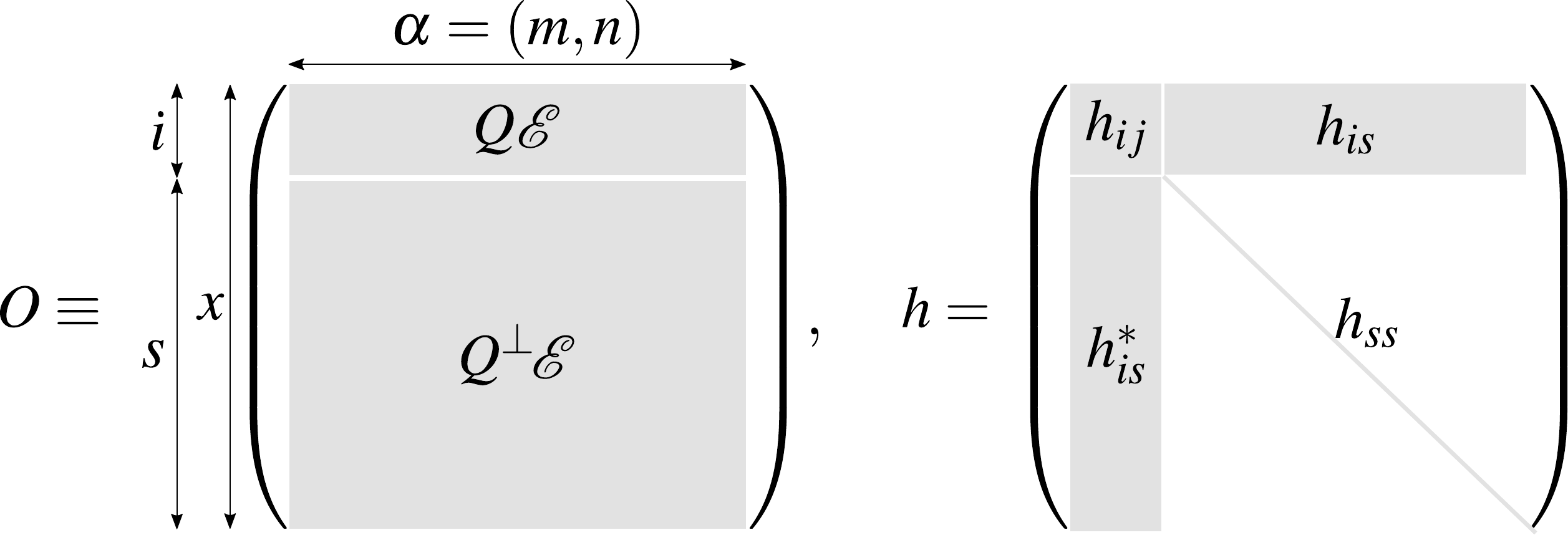}
  \caption{Unitary completion of the time-dependent Matrix $Q(t)$. The matrix
    $Q^{\bot}(t)$ contains a completing set of orthonormal basis vectors in its
    rows. For convenience, the phase factor $\mathcal{E}_{\alpha\alpha'}(t) =
    \delta_{\alpha\alpha'} \exp{-i\epsilon_\alpha t}$ is also absorbed into
    $O(t)$. The generating, Hermitian matrix $h(t)$ (cf.\ Eq.\ \eqref{eq:heff})
    can be assumed to be diagonal in the virtual sector.}
  \label{fig:qmat}
\end{figure}

We start our construction from the Lehmann representation of $G$ as stated in
Eq.\ \eqref{eq:representable}. For our model Hamiltonian \eqref{eq:ham} the
associated one-particle excitation energies $\epsilon_\alpha$ and the
$Q$-matrix are given by Eq.\ \eqref{eq:qmat}.  The self-energy is related to
this representation through Dyson's equation $\Sigma=G_0^{-1}-G^{-1}$. However,
the inverse $G^{-1}$ cannot directly be calculated with Eq.\
\eqref{eq:representable} since $Q(t)$ is not quadratic. As a first step we
block up the matrix $Q(t)$ to a quadratic form.  This is achieved by
interpreting its orthonormal rows (cf.\ Eq.\ \eqref{eq:orthonormal}) as an
incomplete set of basis vectors. $Q(t)$ itself is an incomplete unitary
transform from this viewpoint. We now pick an arbitrary, pairwise orthonormal
completion of this basis to find an unitary transform $O(t)$ that contains
$Q(t)$ in its upper block (cf.\ Fig.\ \ref{fig:qmat}). The next steps of our
discussion will be independent of the particular completion that is chosen. The
only mathematical requirement is that it is as smooth (and thus differentiable)
in the time variable $t$ as $Q(t)$; see Appendix \ref{ap:numerics} for
numerical details on the construction of $O(t)$.

The completed unitary transform $O(t)$ describes additional virtual
orbitals (labeled by the index $s$, see Fig.\ \ref{fig:qmat} and Eq.\
\eqref{eq:heff_eck}). For convenience, we also absorb in the definition of
$O(t)$ the extra factor $\mathcal{E}_{\alpha\alpha'}(t) =
\delta_{\alpha\alpha'} \exp{-i\epsilon_\alpha t}$ that stems from the
non-interacting Green's function $g(\epsilon_\alpha;t,t)$ (cf.\ Eqs.\
\eqref{eq:representable} and \eqref{eq:isolated}). For clarity in the notations
we use the following index convention throughout this paper
\begin{align} 
   \text{physical orbitals:}~~i,j,
  &\quad\text{virtual orbitals:}~~r,s,\nonumber\\ 
   \text{physical {\em or} virtual orbitals:}~~x,y,
  &\quad\text{excitations:}~~\alpha,\alpha'.
\end{align} 
Like every time-dependent unitary transform, $Q(t)$ is generated by an
associated Hermitian matrix. We define
\begin{equation}
	\label{eq:heff}
  h_{xy}(t) = \sum_{\alpha}
              \left[
                i\partial_t O_{x\alpha}(t)
              \right]
              O_{\alpha y}^\dagger(t).
\end{equation}
Indeed, by integration we have
\begin{equation}
  O(t)=\torder\exp{-i\int_0^t h(t')\dt'}O(0) 
\end{equation}
and furthermore $h(t)$ is Hermitian:
\begin{align}
  \label{eq:hermitian}
  h(t) &= [i\partial_t O(t)]O^\dagger(t)
        = i\partial_t [O(t)O^\dagger(t)]-O(t)i\partial_tO^\dagger(t)\nonumber\\
       &= \left(
           [i\partial_t O(t)]O^\dagger(t)
         \right)^\dagger 
        = h^\dagger(t).
\end{align}
We now require the virtual part $h_{ss'}(t)$ to be diagonal and
time-independent, i.e.,  $h_{ss'}(t) = h_{ss}(0)\delta_{ss'}$. To this end we
use our freedom in choosing the completing basis vectors $Q^\bot(t)$ which
allows us to perform the associated unitary transform in the virtual sector
(see Fig.\ \ref{fig:qmat}). With the resulting $h_{xy}(t)$ we define the
single-particle Hamiltonian
$H_\mathrm{eff}(t)$
\begin{equation}
  \label{eq:heff_ham}
  H_\mathrm{eff}(t) = \sum_{xy} h_{xy}(t)\cds{x}\ccs{y},
\end{equation}
which has precisely the form of the effective Hamiltonian stated in Eq.
\eqref{eq:heff_eck}. The requirement of a diagonal virtual sector defines the
effective Hamiltonian uniquely up to rotations in invariant subspaces.

At time $t=0$, the effective medium can be stated in a diagonal form which is
useful for the evaluation of the corresponding one-particle Green's function.
We recall that we required $O(t)$ to be as smooth as $Q(t)$ and take a look at
\begin{equation}
  \label{eq:get_h0}
  [i\partial_t O(t)]_{t=0}=h(0)O(0)=O(0) M,
\end{equation}
where $M=O^\dagger(0) h(0) O(0)$. Eq.\ \eqref{eq:get_h0} implies in particular
that $[i\partial_t Q(t)\mathcal{E}(t)]_{t=0}=Q(0) M$ (cf.\ Fig.\
\ref{fig:qmat}).  However, from Eq.\ \eqref{eq:qmat} one easily evaluates
$[[i\partial_t Q(t)\mathcal{E}(t)]_{i\alpha}]_{t=0} = Q(0)_{i\alpha}
\epsilon_\alpha $ and we can thus identify
$M_{\alpha\alpha'}=\delta_{\alpha\alpha'}\epsilon_\alpha$.  Putting everything
together we find
\begin{equation}
  \label{eq:heff_eq}
  h_{xy}(0)=\sum_\alpha
  O_{x\alpha}(0)\epsilon_\alpha O^*_{y\alpha}(0).
\end{equation}

We require that the effective medium is initially in thermal equilibrium with
the same inverse temperature $\beta$ and the same chemical potential $\mu$ as
the physical system. The associated one-particle Green's function of the medium
is defined as
\begin{equation}
  F_{xy}(t,t') = -i\est{\tcorder\cc{x}(t)\cd{y}(t')}_{H_\mathrm{eff}}.
\end{equation}
Recalling the diagonal form of the effective medium at $t=0$ (cf.\ Eq.\
\ref{eq:heff_eq}) and using that the effective Hamiltonian \eqref{eq:heff_ham}
is non-interacting, we can easily rewrite this expression into
\begin{equation}
  F_{xy}(t,t') = i\sum_{\alpha}O_{x\alpha}(t)[f(\epsilon_\alpha) -
  \Theta_C(t,t')] O^*_{y \alpha}(t).
\end{equation}
The physical sector of $F$ is by construction identical with the
Lehmann representation of $G$:
\begin{equation}
  \label{eq:F_eq_G}
  F_{ij}(t,t') 
  = \sum_{\alpha}Q_{i\alpha}(t) g(\epsilon_\alpha; t,t')Q_{j\alpha}^*(t')
  = G_{ij}(t,t').
\end{equation}
$F$ encodes the full information on the one-particle excitations of the system
defined by the Hamiltonian \eqref{eq:ham}. Eq.\ \eqref{eq:F_eq_G} further
stresses the fact that in principle any (sufficiently smooth) completion of
$Q(t)$ to a unitary transform $O(t)$ leads to a valid effective Hamiltonian.
The physical sectors of $O(t)$ and $h(t)$ remain independent of its choice. The
virtual sectors, on the other hand, are affected and only the special choice of
$O(t)$ (cf.\ the discussion above and below Eq.\ \eqref{eq:heff_ham})
guarantees a diagonal form of the effective medium.

Having found an effective, non-interacting model that reproduces the correct
Green's function, it remains to link this back to the self-energy.
The time-non-local (correlated) part $\Sigma^\mathrm{C}_{ij}(t,t')$ follows by
tracing out the virtual orbitals.  This procedure is straightforward as they
are all non-interacting and we can use, e.g., a cavity-like ansatz \cite{cg:13}
or an equation of motion based approach. \cite{eck:01.14} This results in a
hybridization-like function 
\begin{equation}
  \label{eq:sigma_corr}
  \Sigma^\mathrm{C}_{ij}(t,t')\equiv\sum_s h_{is}(t)g(h_{ss};t,t')h^*_{js}(t')
\end{equation}
that encodes the influence of the virtual sites on the physical sector. The
Green's function at the physical orbitals is then obtained from a Dyson-like
equation
\begin{align}
  \label{eq:dysonlike}
	F_{ij}(t,t') = \left[
                 \frac{1}{F_0^{-1} - \Sigma^\mathrm{C}}
                 \right]_{ij}(t,t'),
\end{align}
where
\begin{align}
  [F_0^{-1}]_{ij}(t,t')=[i\partial_t - h_{ij}(t)]\delta_C(t,t'),
\end{align}
with $\delta_C(t,t')=\partial_t \Theta_C(t,t')$ as the contour delta function.

To make the final connection to the self-energy we evaluate the physical sector
of $h$. With
\begin{align}
  i\partial_t &Q_{i(m,n)}(t)e^{-i\epsilon_{(m,n)} t}
  = z_{(m,n)}\lwf{m}[\cc{i}(t),\hat{H}(t)]\rwf{n}\nonumber\\
	&=  \sum_{j}\left(T_{ij}(t)-\mu\delta_{ij}\right)Q_{j(m,n)}(t)\nonumber\\
  &\phantom{=}~~+\sum_{ji'j'}U_{ii'jj'}(t)z_{(m,n)}\lwf{m} 
  \cd{i'}(t)\cc{j'}(t)\cc{j}(t)\rwf{n}
\end{align}
we obtain
\begin{align}
  \label{eq:heff_phys}
  h_{ij}(t) &= T_{ij}(t)-\delta_{ij}\mu+\Sigma^\mathrm{HF}_{ij}(t),\nonumber\\
  \Sigma^\mathrm{HF}_{ij}(t) &\equiv 2\sum_{i'j'}U_{ii'jj'}(t)
  \est{\tcorder\cd{i'}(t)\cc{j'}(t)}_{H_\mathrm{eff}}.
\end{align}
At the physical orbitals the effective Hamiltonian is thus determined by the
Hartree-Fock Hamiltonian. By comparison of Eq.\ \eqref{eq:dysonlike} with
the Dyson equation
\begin{equation}
	G_{ij}(t,t') = \left[
                    \frac{1}{G_0^{-1} - \Sigma}
                  \right]_{ij}(t,t'),
\end{equation}
where
\begin{equation}        
  [G_0^{-1}]_{ij}(t,t') 
               = \left[
                    i\partial_t - (T_{ij}(t)-\mu\delta_{ij})
                  \right]\delta_C(t,t'),
\end{equation}
we finally identify
\begin{equation}
  \label{eq:sigma_full}
  \Sigma_{ij}(t,t') = \delta_C(t,t')\Sigma^\mathrm{HF}_{ij}(t) 
  +\Sigma^\mathrm{C}_{ij}(t,t'),
\end{equation}
concluding our construction of the self-energy. Let us stress that with Eqs.
\eqref{eq:heff_ham}, \eqref{eq:sigma_corr} and \eqref{eq:heff_phys} we now have
an {\em explicit} recipe to construct the Lehmann representation of the
self-energy. This representation is further unique as follows from the
uniqueness of the corresponding effective Hamiltonian (cf.\ the discussion
above and below Eq.\ \eqref{eq:heff_ham}).

\subsection{Useful properties}
\label{sec:use_prop}
With the Hamiltonian of the effective medium, Eq.\ \eqref{eq:heff_ham}, at
hand, a number of useful properties follow immediately:

\subsubsection{Positive spectral weight}
By taking a look at the Matsubara branch only, one can link the Lehmann
representation of the self-energy to the positive definiteness of its
equilibrium spectral function.  With $\Sigma^\mathrm{M}(\tau-\tau') \equiv
-i\,\Sigma(-i\tau,-i\tau')$ we can perform the usual Fourier transform from
imaginary time to Matsubara frequencies and then find the analytical
continuation $\Sigma^\mathrm{M}(\omega)$ to the complex-frequency plane (see
for example Ref.  \onlinecite{cg:13}).  The spectral function is defined as 
\begin{align}
  C^\Sigma_{ij}(\omega) = \frac{i}{2\pi}[\Sigma_{ij}^\mathrm{M}(\omega + i0)-
  \Sigma_{ij}^\mathrm{M}(\omega - i0)]
\end{align}
for real $\omega$. This can explicitly be calculated from the parameters of the
effective Hamiltonian. One finds:
\begin{align}
  C^\Sigma_{ij}(\omega) =\sum_{s} h_{is}(0)h_{js}^*(0) \delta(\omega - h_{ss}),
\end{align}
where $\delta(\omega)$ is the Dirac delta function. The positive definiteness
for every $\omega$ is immediately evident.

\subsubsection{Higher-order correlation functions}
The self-energy and its time derivatives can be used to calculate certain
expectation values of higher order. Prominent examples include the interaction
energy or the local double occupation. Their calculation is based on the 
evaluation of contour integrals of the form $\int_C\dt'
\Sigma(t,t')G(t',t)$.  By comparing the equations of motion for
$G_{ij}(t,t')$ and $F_{xy}(t,t')$ one readily finds the identity
\begin{align}
  \label{eq:tomarkovian}
  \int_C\mathrm{d}\tilde{t}\sum_{j}
  \Sigma_{ij}(t,\tilde{t})G_{ji'}(\tilde{t},t')=\sum_j &[h_{ij}(t)-T_{ij}(t)]
  F_{ji'}(t,t')\nonumber\\
  &+\sum_s h_{is}(t) F_{si'}(t,t').
\end{align}
This is a remarkable relation as the contour integration can be avoided in
favor of a simple matrix multiplication.

\subsubsection{Quantum quenches}
A convenient tool to drive quantum systems out of equilibrium is given by the
so-called quantum quenches. Here, one (or more) parameters of the system are
changed suddenly. This sudden change reflects itself as a discontinuous time
dependence of the effective Hamiltonian: Assume that the system is subjected to
a quench at time $t=0$, so that $H_\mathrm{ini} \rightarrow H_\mathrm{final} =
\mathrm{const}$. Initially the system is in thermal equilibrium and the
effective Hamiltonian is given by Eq.\ \eqref{eq:heff_eq}, where
$\epsilon_\alpha$ are the excitations energies of $H_\mathrm{ini}$.  The
$O$-matrix is continuous at $t=0$ despite the quantum quench (it only depends
on $\cc{i}(t)$, cf.\ Eq.\ \eqref{eq:qmat}). Its time derivative, however, is
not and thus $h(t)$ jumps from $h(0)$ to
\begin{equation}
  h_{ij}(0^+)=\sum_\alpha [i\partial_t O_{i\alpha}(t)]_{t=0^+}
  O^*_{j\alpha}(0).
\end{equation}
After this jump, the effective Hamiltonian will in general not be constant for
times $t>0$, i.e., $h(t) \neq h(0^+)$.

\section{Application to cluster-perturbation theory}
\label{sec:appl_cpt}
The simplest numerical application of our formalism is given by
cluster-perturbation theory\cite{gros:93, sen:00, sen:02, bal:05.11, knap:11,
  jur:13} (CPT). The idea of CPT is to split the system into small clusters
which can be treated by means of exact-diagonalization techniques. The cluster
self-energies are then used as approximate input for the Dyson equation
\eqref{eq:dyson_volterra} to obtain the CPT Green's function. The same concept
is part of more powerful approaches like DMFT \cite{voll:01.89, kot:03.92,
  sch:02, fre:06} or self-energy functional theory \cite{po:03,hof:13} where
the CPT Green's function is self-consistently or variationally linked to the
self-energy of a reference system. The following construction of an effective
Hamiltonian for CPT applies to such techniques as well.

\subsection{Cluster-perturbation theory (CPT)}
\label{sec:cpt_recalled}

\begin{figure}[b]
  \includegraphics[width=0.2\textwidth]{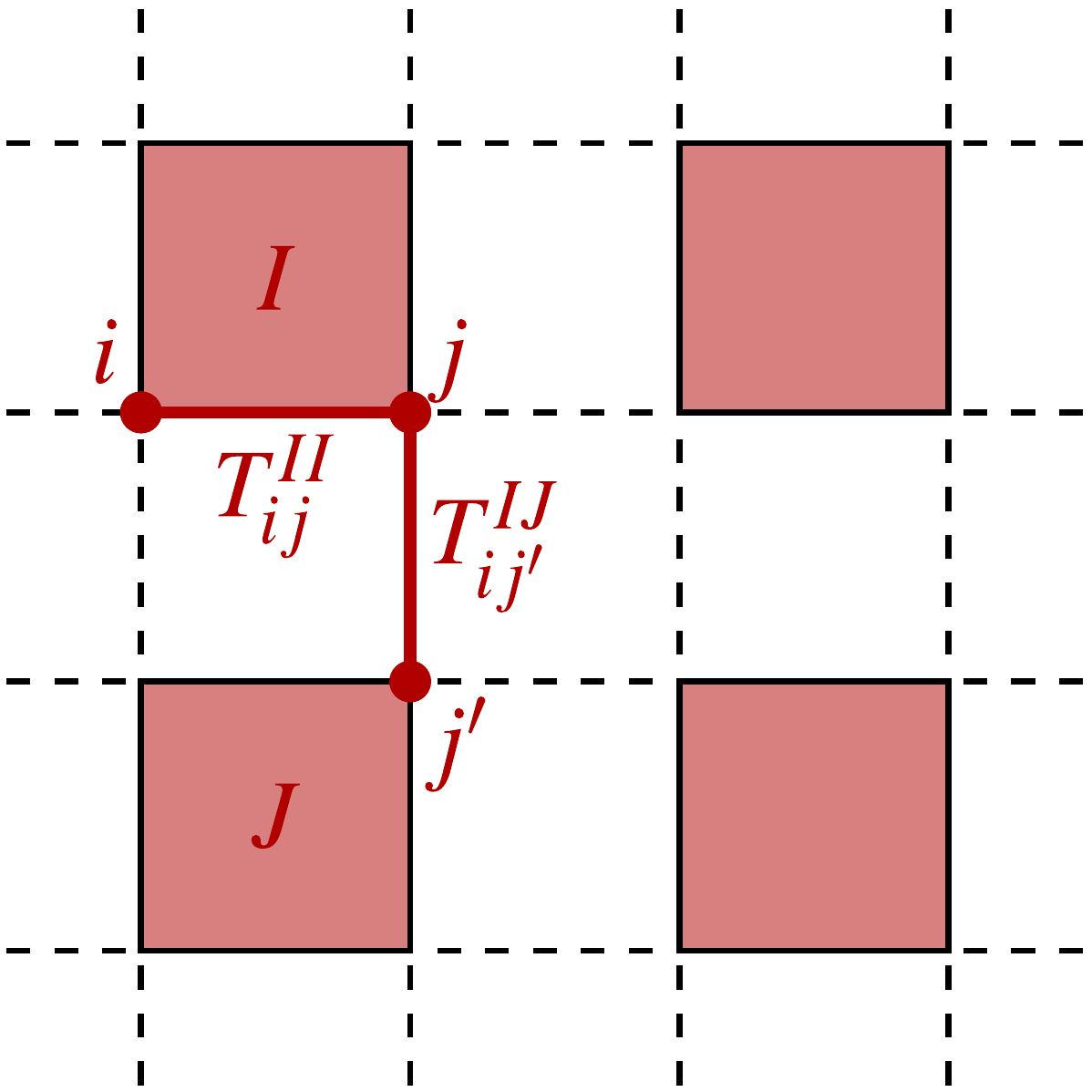}
  \caption{Illustration of the partitioning of an infinite, two-dimensional
    square lattice into $2\times2$ clusters. The sites $i,j$ lie within the
    same cluster $I$, $j'$ belongs to a different cluster $J$. The cluster
    diagonal part of the hopping matrix $T^{II}_{ij}$ describes the
    intra-cluster, the cluster off-diagonal part $T^{IJ}_{jj'}$ ($I\neq J$) the
    inter-cluster-hopping.}
  \label{fig:cpt}
\end{figure}

From now on we restrict ourselves to the fermionic Hubbard model. The locality
of its interaction term allows us to cast its Hamiltonian into the following
form:
\begin{align}
  \label{eq:cpt_hubbard}
      H(t)=&\sum_{I}\underbrace{
              \left[
                \sum_{ij\sigma} [T^{II}_{ij\sigma}(t)-\mu\delta_{ij}]
                  \cds{Ii\sigma}\ccs{Ij\sigma}
                + U(t)\sum_i n_{Ii\uparrow}n_{Ii\downarrow}
              \right]}_{\text{cluster system~}H_I} \nonumber\\
     &\,\, +\underbrace{
                \sum_{I\neq J}\sum_{ij\sigma}
                T^{IJ}_{ij\sigma}(t)\cds{Ii\sigma}\ccs{Jj\sigma}
              }_\text{inter-cluster hopping}.
\end{align}
Here, the indices $I,J$ label the cluster systems, while the indices $i,j$ run
over the sites within a cluster only (see Fig.\ \ref{fig:cpt}). Of course,
this is fully equivalent with the usual form of the Hubbard model which is
re-obtained by combining $(I,i)$ to a superindex, i.e., $(I,i)\rightarrow i$.
The operator $n_{Ii\sigma}=\cds{Ii\sigma}\ccs{Ii\sigma}$ measures the particle
density with spin projection $\sigma =\uparrow,\downarrow$. The Green's
function of the isolated cluster $I$ with intra-cluster Hamiltonian $H_I$ is
\begin{equation}
  G^I_{ij\sigma}(t,t')=-i\est{\tcorder \cc{Ii\sigma}(t)\cd{Ij\sigma}(t')}_{H_I},
\end{equation}
so that
\begin{equation}
  [G^I]^{-1}_{ij\sigma}(t,t')=
  [i\partial_t- (T^{II}_{ij\sigma}(t)-\mu\delta_{ij})]
  \delta_C(t,t') - \Sigma^I_{ij\sigma}(t,t'),
\end{equation}
where $\Sigma^I$ denotes the corresponding self-energy. We further define
\begin{equation}
  G'=\begin{pmatrix}
          G^1 & 0 & \cdots\\
          0 & G^2  & \cdots\\
          \vdots & \vdots & \ddots
        \end{pmatrix},\quad
  \Sigma'=\begin{pmatrix}
          \Sigma^1 & 0 & \cdots\\
          0 & \Sigma^2  & \cdots\\
          \vdots & \vdots & \ddots
        \end{pmatrix}.\quad
\end{equation}
With the inter-cluster ($\mathrm{ic}$) hopping $[T^\mathrm{ic}]^{IJ}_{ij\sigma}
= (1-\delta_{IJ}) T^{IJ}_{ij\sigma}$ the CPT Green's function is defined as
\begin{equation}
	\label{eq:cptdyson}
  G^\mathrm{CPT}\equiv \frac{1}{(G')^{-1}-T^\mathrm{ic}}
                = \frac{1}{G_0^{-1}-\Sigma'},
\end{equation}
where $[G_0^{-1}]^{IJ}_{ij\sigma}(t,t')=[i\partial_t -(T^{IJ}_{ij\sigma}(t) -
\mu \delta_{IJ}\delta_{ij})]\delta_C(t,t')$. The definition of $G^\mathrm{CPT}$
reveals that CPT becomes exact in the limit of vanishing interaction. We then
have $\Sigma'=0$ and thus $G^\mathrm{CPT} = [G_0^{-1}]^{-1}=G_0$. Solving Eq.
\eqref{eq:cptdyson} in case of non-vanishing $\Sigma'$, on the other hand,
requires the solution of a Dyson equation. This brings us back to our original
problem.

\subsection{Application of the Lehmann representation for the self-energy}
\label{sec:appl_lehmann}
Using our results from Sec.\ \ref{sec:lsig} we can avoid the solution of the
Dyson equation and rather decompose the self-energies of the isolated clusters
into their Lehmann representations:
\begin{align}
  \Sigma^I_{ij\sigma}(t,t')=&\delta_C(t,t')
  [\Sigma^\mathrm{HF}]^I_{ij\sigma}(t)\nonumber\\
  &+\sum_{s}h^{I}_{is\sigma}(t)g(h^I_{ss\sigma};t,t')[h^{I}]^*_{js\sigma}(t').
\end{align}
Here, $h^I(t)$ are the parameters of the effective medium corresponding to the
$I$-th cluster. We define 
\begin{equation}
  h'(t)=\begin{pmatrix}
    h^1(t) & 0 & \cdots\\
    0 & h^2(t)  & \cdots\\
    \vdots & \vdots & \ddots
\end{pmatrix}.
\end{equation}
It is now straightforward to realize that the inclusion of the inter-cluster
hopping by means of Eq.\ \eqref{eq:cptdyson} is completely trivial in this
language. Namely,
\begin{equation}
  h^\mathrm{CPT}(t)=h'(t) + T^\mathrm{ic}(t).
  \label{eq:cpt_hamiltonian}
\end{equation}
With
\begin{equation}
  H^\mathrm{CPT}(t)=\sum_{IJ}\sum_{xy\sigma}[h^\mathrm{CPT}]^{IJ}_{xy\sigma}
                      \cds{Ix\sigma}\ccs{Jy\sigma},
\end{equation}
we then have
\begin{equation}
  [G^\mathrm{CPT}]^{IJ}_{ij\sigma}(t,t')=-i\est{\tcorder
    \cc{Ii\sigma}(t)\cd{Jj\sigma}(t')}_{H^\mathrm{CPT}}.
\end{equation}
While this is an easy and intuitive description, we remark that $h'(t)$
includes virtual orbitals. The inter-cluster hopping $T^\mathrm{ic}(t)$, on the
other hand, is defined solely in the physical sector and has to be blocked
up accordingly ($[T^\mathrm{ic}(t)]^{IJ}_{rs\sigma} =
[T^\mathrm{ic}(t)]^{IJ}_{is\sigma} = [T^\mathrm{ic}(t)]^{IJ}_{rs\sigma}=0$).

As an important observable we briefly discuss the calculation of the total
energy within CPT. While the kinetic energy follows straightforwardly from the
one-particle density matrix as $E_\mathrm{kin}(t) = -i\sum_{IJ} \sum_{ij\sigma}
T_{ij\sigma}^{IJ} G_{ij\sigma}^{IJ}(t,t^+)$, the interaction energy can only be
accessed indirectly through the self-energy. It is given by
\begin{align} 
  \label{eq:cpteint}
  E_\mathrm{int}(t)=-i\sum_{ij}\int_C \dt_1 
    \Sigma'_{ij\sigma}(t,t_1)G^\mathrm{CPT}_{ji\sigma}(t_1,t^+).
\end{align} 
The evaluation of this contour-integral in Eq.\ \eqref{eq:cpteint} is
straightforward within our formalism by using Eq.\ \eqref{eq:tomarkovian}.

\section{Numerical results}
\label{sec:numerics}
\subsection{Prethermalization}
\label{sec:numerics_intro}
The study of real-time dynamics initiated by an interaction quench in the
Hubbard model has attracted much attention recently. \cite{col:07, tro:12,
  sch:12, eck:03.10, eck:07.09, moe:08, stk:13}  Here, the system is prepared
in a thermal (usually non-interacting) initial state and then, after a sudden
change of the interaction parameter $U$, evolves in time as prescribed by the
interacting Hamiltonian. While the setup is apparently simple, the search for
universal properties of the time evolution remains notoriously difficult due to
the non-integrability of the Hubbard model in two and higher dimensions. Apart
from the general assumption that non-integrable models feature thermalization
and thus lose memory of the initial state in the long-time
limit,\cite{rig:04.08} only the time evolution after quenches to a weak, finite
Hubbard $U$ seems to be well understood so far.  Here, it could be shown by
means of weak-coupling perturbation theory \cite{moe:08, moe:09, moe:10,
  stk:13} that observables initially relax to non-thermal, quasistationary
values (the system prethermalizes) before the significantly slower relaxation
towards the thermal values sets in.
 
It was later worked out\cite{eck:08.11} that the mechanism which traps the
system in a quasi-stationary prethermal state is quite similar to the mechanism
that hinders non-interacting systems from thermalizing. In the latter case the
integrability of the Hamiltonian leads to a large number of constants of motion
that highly constrain the dynamics of the system. In case of weakly interacting
systems it is the proximity to the integrable point that introduces approximate
constants of motion and hinders relaxation beyond the prethermalization
plateau on short timescales $t\lesssim T/U^2$ (here, $T$ is the
nearest-neighbor hopping).  Relaxation towards the thermal average is delayed
until later times ($t\gtrsim T^3/U^{4}$).

As a proof of concept of our formalism we use nonequilibrium CPT to investigate
the short- and long-time dynamics of an inhomogeneous initial state after an
interaction quench in the Hubbard model. In particular we will study if and to
what extent the CPT is able to describe prethermalization and the subsequent
relaxation to a thermal state.

\subsection{Setup}
We consider the Hubbard model at zero temperature ($\beta\rightarrow\infty$)
and half-filling ($\mu=U/2$) on a square lattice of $L=10\times10$ sites with
periodic boundary conditions.  Cluster indices run over $I,J \in \{ 0, 1,
\dots, 24\}$ and $i,j \in \{0, 1, 2, 3\}$, so that the system is cut into $25$
clusters of size $2\times 2$. The hopping is restricted to nearest neighbors
and we set $T=1$ to fix energy and time units.  Translational invariance of the
initial state is broken by applying a local magnetic field of strength $B$ to
an arbitrarily chosen ``impurity site'' (here, site $0$ in \makebox{cluster
  $0$}):
\begin{equation}
  \label{eq:cpt_hopping}
  T_{ij\sigma}^{IJ}(t)
  =
  \delta_{\langle(I,i),(J,j)\rangle}  T 
  -
  z_{\sigma} \delta_{I,J}\delta_{i,j} \delta_{I,0}\delta_{i,0} \, B(t) \; ,
\end{equation}
where $\delta_{\langle...\rangle}$ is non-zero and unity for nearest neighbors
only and where $z_{\uparrow} = +1$ and $z_{\downarrow} = -1$.  Initially, the
magnetic field is switched on with strength $B(0)=10$ to induce a (nearly)
fully polarized magnetic moment on the impurity site and then switched off for
times $t>0$:
\begin{equation}
  B(t)= B(0) (1-\Theta(t)).
\end{equation}
Here, $\Theta(t)$ is the Heaviside step function.  Furthermore, the interaction
$U(t)$ is switched off initially and then switched on to a non-zero value
$U_\mathrm{fin}$
\begin{align}
  \label{eq:cpt_interaction}
  U(t)=U_\mathrm{fin}\,\Theta(t).
\end{align}
Hence, in the quantum quench considered here, two parameters are changed
simultaneously.  The initial Hamiltonian $H_\mathrm{ini}$ features no
interactions but is inhomogeneous due to the local magnetic field, the final
Hamiltonian $H_\mathrm{fin}$ is translationally invariant due to the absence of
the magnetic field but has a finite interaction $U_\mathrm{fin}>0$. 

To apply nonequilibrium CPT, we use exact diagonalization to solve the $25$
independent cluster problems and to construct the Hamiltonian of the effective
medium (for details on the numerical implementation see Appendix
\ref{ap:numerics}). Finally, Eq.\ \eqref{eq:cpt_hamiltonian} is used to account
for the inter-cluster hopping. The number of non-zero elements of a cluster's
$Q$-matrix and therefore the computational effort of our approach increases
quadratically with the number of active states in the density matrix
$\rho_\text{cluster} = \sum_m \exp{-\beta E_m} \rwf{m}\lwf{m}$
($H_\text{cluster} \rwf{m}=E_m\rwf{m}$), i.e., states that contribute with a 
significant weight $\exp{-\beta E_m}$ to thermal averages.
For convenience we have therefore chosen a zero-temperature initial state and 
consider a weak interaction $U=10^{-4}$ to lift the ground-state degeneracy
present in the non-interacting system (denoted as $U=0^{+}$ in
the following). The effective Hamiltonian $h^{I}(t)$ for each cluster is then
of size $48\times 48$ and the final CPT Hamiltonian of size $1200\times 1200$.
Exploiting its sparse form we are able to perform $1,000,000$ time steps with
$\Delta t = 0.01$ to reach a maximal time $t_\text{max}=10^{4}$ with modest
computational effort. For comparison we note that prior studies based on the
nonequilibrium CPT, e.g.\ Refs.\ \onlinecite{bal:05.11,jur:13}, have been limited 
to $t_\text{max} = 10$--$20$ inverse hoppings.

\begin{figure}
  \includegraphics[width=\linewidth]{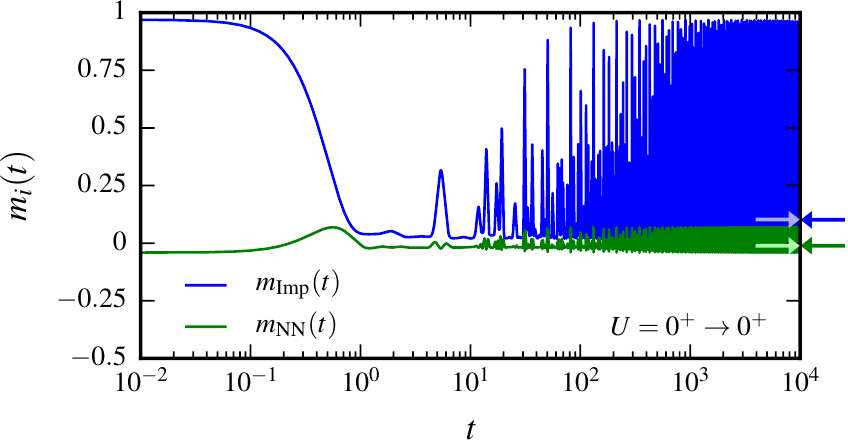}
  \caption{(Color online) Time evolution of the local magnetic moment at the
    impurity ($m_\mathrm{Imp}(t)$, blue line) and its nearest neighbors
    ($m_\mathrm{NN}(t)$, green line). The dark-blue (dark-green) arrow, which
    is pointing from right to left, indicates the long time average of the blue
    (green) curve. The light-blue (light-green) arrow, which is pointing from
    left to right, indicates the analytical average \eqref{eq:ana_avg}. The
    long time average was taken over $500,000$ data points in the interval
    $[0.5\times 10^4, 10^4]$.}
  \label{fig:non_int}
\end{figure}

The partitioning of the lattice into $2\times 2$ clusters by CPT breaks
rotational and reflection symmetries of the original problem. These are
restored by averaging the resulting one-particle density matrix over the $4$
possible ways to cut the lattice into $2\times 2$ clusters. In the following we
will show results for the time evolution of the local magnetic moment
$m_i(t)=n_{i\uparrow}(t)-n_{i\downarrow}(t)$ at the impurity
($m_\mathrm{Imp}(t)$) and at its nearest neighbors ($m_\mathrm{NN}(t)$).  Only
the latter are affected by the averaging. It restores the equivalence of
nearest neighbors that lie in the same and nearest neighbors that lie in a
neighboring cluster of the impurity. The extensive quantities total energy
$E_\mathrm{tot}(t) = E_\mathrm{kin}(t) + E_\mathrm{int}(t)$ (cf.\  Eq.\
\eqref{eq:cpteint} and preceding discussion) and total magnetization
$M(t)=\sum_i m_i(t)$ are both unaffected by the averaging.

\begin{figure*}
  \includegraphics[width=\textwidth]{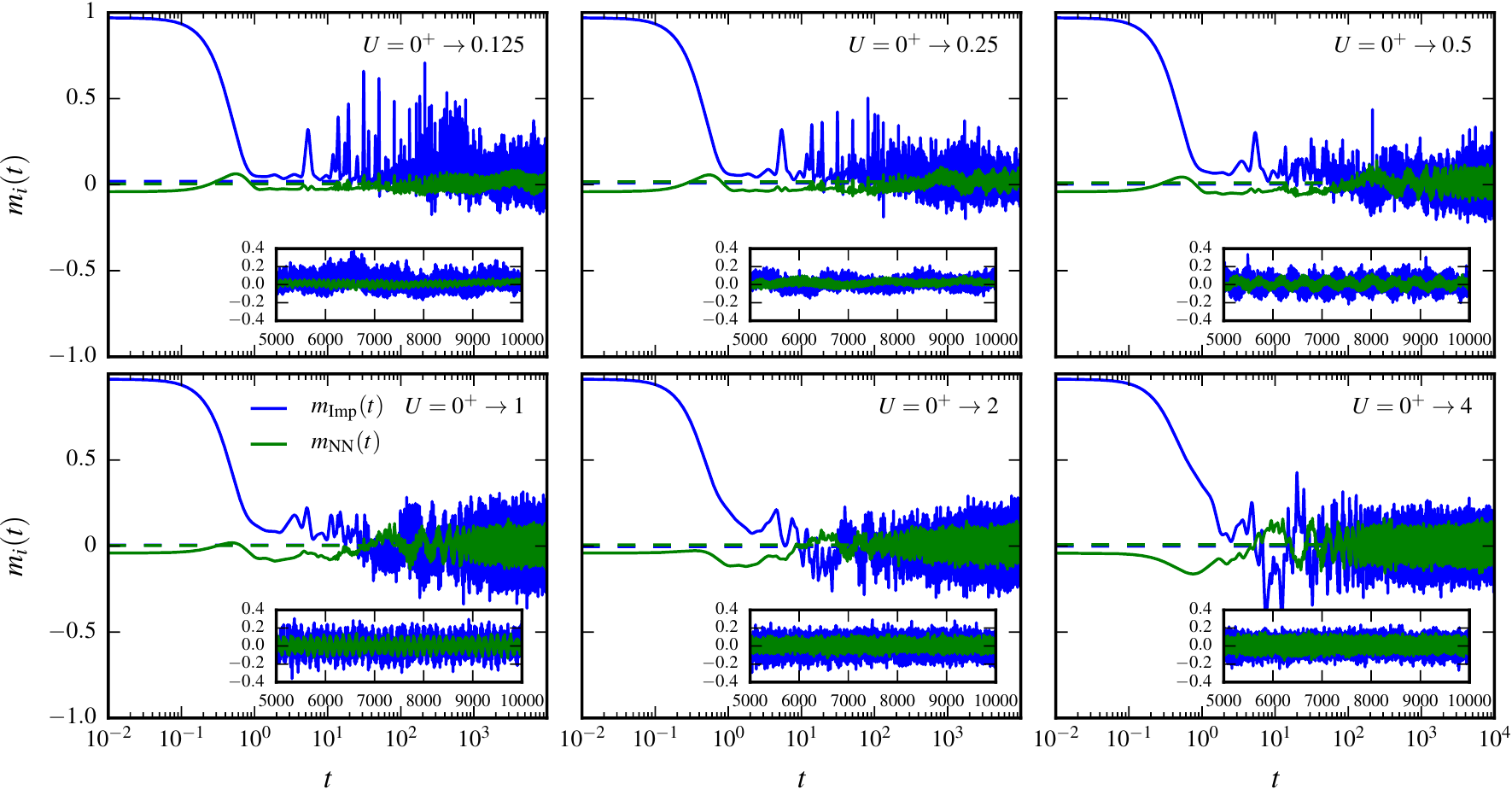}
  \caption{(Color online) CPT results for the time evolution of the local
    magnetic moment at the impurity ($m_\mathrm{Imp}(t)$, blue line) and at its
    nearest neighbors ($m_\mathrm{NN}(t)$, green line) for quenches from the
    limit of vanishing interaction $U=0^+$ (numerically implemented by setting
    $U=0.0001$) to finite $U_\mathrm{fin}$. In the insets the long-time
    behavior ($t\in [5\times 10^3, 10^4]$) is plotted on a linear scale. The
    interval consists of $500,000$ data points and was also used to calculate
    the long-time average (straight dashed lines). In total 1,000,000 time
    steps were performed with $\Delta t=0.01$ on a $L=10\times 10$ lattice (cut
    into $25$ clusters of size $2\times 2$ by CPT).  }
  \label{fig:uquench}
\end{figure*}

The initial state is the same for all quenches discussed in the following. We
find a polarization of $m_\mathrm{Imp}(0)\approx 0.97$ at the impurity
which is partially screened (e.g., $m_\mathrm{NN}(0)=-0.04$) so that the total
magnetization amounts to $M(0)=\sum_{i} m_i(0)\approx 0.70$.

\subsection{Noninteracting case}
We first discuss the non-interacting case, i.e., a purely magnetic quench where
$U_\mathrm{fin}=0^{+}$. Here, CPT predicts the exact time evolution (cf.\ the
discussion below Eq.\ \eqref{eq:cptdyson}) since the cluster self-energies
$\Sigma^{I}$ vanish. Our results are shown in Fig.  \ref{fig:non_int}. For
short times ($t\in [10^{-2}, 4\times 10^0]$) the local magnetic moment at the
impurity $m_\mathrm{Imp}(t)$ (blue line) decays to a value slightly above zero.
Subsequently ($t\in [4\times 10^0, 10^4]$) the dynamics is governed by
collapse-and-revival oscillations caused by the finite system size.  In
particular we find that $m_\mathrm{Imp}(t)$ returns arbitrarily close to its
initial value for large times. This is readily understood from the fact that
the system's dynamics is governed by the one-particle propagator $\exp{-i
  T_\mathrm{fin} t}$ where $T_\mathrm{fin}$ denotes the final hopping matrix
(i.e., after the quench).  $T_\mathrm{fin}$ involves only a small number of
different one-particle energy levels and thus $U(t,0)$ returns arbitrarily
close to the identity matrix over time.

For the non-interacting system it is possible to directly access the long-time
average of the one-particle density matrix. One finds
\begin{align}
  \label{eq:ana_avg}
  \rho_{ij\sigma}^\mathrm{avg}&=\lim_{t_\mathrm{max}\rightarrow
    \infty}\frac{1}{t_\mathrm{max}}\int_0^{t_\mathrm{max}}\dt
  \est{\cd{i\sigma}(t) \cc{j\sigma}(t)}\nonumber\\
  &=\frac{1}{L}\sum_{\vec{k}\vec{k'}}\delta_{\varepsilon_{\vec{k}},
    \varepsilon_{\vec{k}'}}e^{i(\vec{k}\cdot\vec{R_i}-\vec{k}'\cdot\vec{R_j})}
  \est{\cd{\vec{k}\sigma}(0)\cc{\vec{k}'\sigma}(0)},
\end{align}
where we used that $H_\mathrm{fin}$ can be diagonalized by a Fourier
transformation involving the reciprocal lattice vectors $\vec{k}$ ($\vec{R}_i$
denotes the lattice vector to site $i$). We then have $H_\mathrm{fin} =
\sum_{\vec{k}\sigma} \varepsilon_{\vec{k}} \cd{\vec{k}\sigma}
\cc{\vec{k}\sigma}$ and $\cc{i\sigma}(t) = \frac{1}{\sqrt{L}} \sum_{\vec{k}}
e^{-i\vec{k} \cdot \vec{R_i}} e^{-i\varepsilon_{\vec{k}} t}
\cc{\vec{k}\sigma}(0)$, where $L$ is the system size. In Fig.\
\ref{fig:non_int} this prediction is compared with the numerical time average
and indeed shows perfect agreement. It is interesting to note that for
non-degenerate energy levels $\varepsilon_k$ one would have
$\rho^\mathrm{avg}_{ii\sigma} = N_\sigma/L$, where $N_\sigma$ is the total
number of particles with spin $\sigma$, and therefore $m^\mathrm{avg}_i =
M(0)/L$.  We conclude that degeneracy of energy levels is required to find
memory of the initial state encoded in the average local magnetic moments
$m^\mathrm{avg}_i$.

\subsection{Quenches to finite $U_\mathrm{fin}$}
For finite $U_\mathrm{fin}$ CPT becomes an approximation and it is a priori
unclear what kind of phenomena it is able to describe. In Fig.
\ref{fig:uquench} we show the long-time evolution for quenches to different
$U_\mathrm{fin}$. For weak $U_\text{fin} \lesssim 0.5$ we find a
(prethermalization-like) separation into two different time scales. Initially
the time evolution qualitatively follows the non-interacting case, i.e., we see
a fast decay of the local moment at the impurity site (blue line) followed by a
quasi-stationary region of collapse-and-revival oscillations. For larger times
these oscillations decay and the system relaxes into a state characterized by
quasi-periodic fluctuations around its long-time average (dashed blue line)
which are driven by different frequencies. Taking a look at the
$U_\mathrm{fin}$ dependence of the dynamics we notice that the region of
collapse-and-revival oscillations shrinks with increasing $U_\mathrm{fin}$ and
finally vanishes for $U_\mathrm{fin}\gtrsim 1$. The system then directly
relaxes into a state with fluctuations around its long-time average. 

For comparison, also the magnetic moment at the neighbouring sites
$m_\mathrm{NN}(t)$ is plotted. While its dynamics for short times must
naturally be different from $m_\mathrm{Imp}(t)$ due to the inhomogeneous
initial state, we would expect a qualitative agreement in the long-time limit
if the system thermalizes. However, this is not the case. There remains a clear
difference in the amplitude of the fluctuations around the long-time average up
to the largest simulated times. Hence we conclude that the system still keeps
memory of the initial state and thus does not thermalize.

Having in mind the general discussion on prethermalization in Sec.\
\ref{sec:numerics_intro}, one can give an intuitive interpretation of these
observations based on the effective-medium approach: While the non-interacting
system is isolated and its dynamics is constrained through many constants of
motion, there is a large number of virtual orbitals coupled to the system in
the interacting case. These virtual orbitals act like a surrounding bath. For
weak $U_\mathrm{fin}$ the virtual orbitals are only weakly coupled to the
system and their influence is delayed to large times, while initially the
dynamics is constrained similar to the non-interacting case. For strong
$U_\mathrm{fin}$, on the other hand, the coupling is strong and affects the
dynamics of the system considerably.  However, the number of virtual sites is
still too small to allow for a complete dissipation of the information on the
initial state into the bath. Therefore, a thermalized state is not reached.
For an exact calculation the number of virtual sites would scale exponentially
in system size. For CPT, on the other hand, it scales exponentially only in
cluster size but linearly in the number of clusters and thus in the system
size.  Memory of the initial state is therefore retained within the
one-particle density matrix and leaves its traces in the magnetic moments as
seen in our calculations.

\subsection{Violation of conservation laws}
CPT as an approximation lacks any kind of self-consistency and is thus unable
to respect the fundamental continuity equations and their corresponding
conservation laws. \cite{hof:13} Therefore, one has to expect a violation of
energy- or particle-number conservation, for example. Furthermore, in contrast
to the equilibrium case where CPT interpolates between the exact limits $U=0$
and $T=0$, it yields exact results only for quenches to $U_\text{fin}=0$.  The
dynamics after a quench to the atomic limit $T_\text{fin}=0$ (with finite
$U_\mathrm{fin}>0$) cannot be described exactly due to the non-local entanglement
of the initial state. We thus generally expect that the quality of the CPT
results degrades with increasing interaction strength.  

\begin{figure}[t]
  \includegraphics[width=\linewidth]{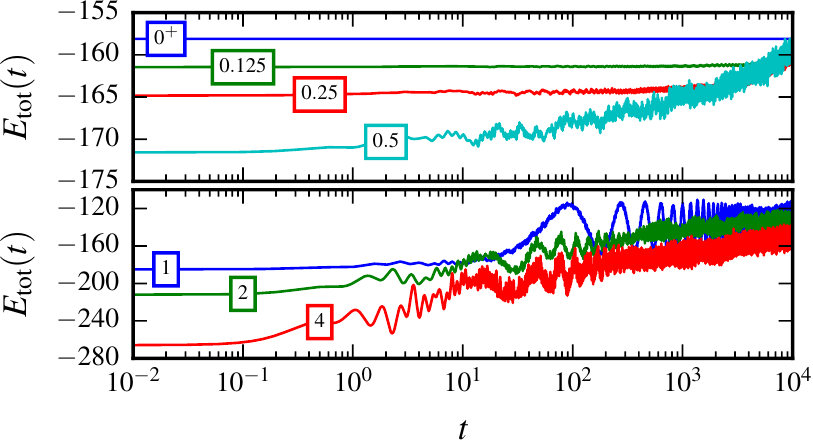}
  \caption{(Color online) Violation of energy conservation by CPT. The numbers
    indicate the respective value of $U_\mathrm{fin}$. Energy conservation is 
    respected for $U_\mathrm{fin}=0^+$ where CPT is exact (blue line). An
    increasingly significant violation of energy conservation is seen for larger
    $U_\mathrm{fin}$.}
  \label{fig:energy}
\end{figure}

\begin{figure}[t]
    \includegraphics[width=\linewidth]{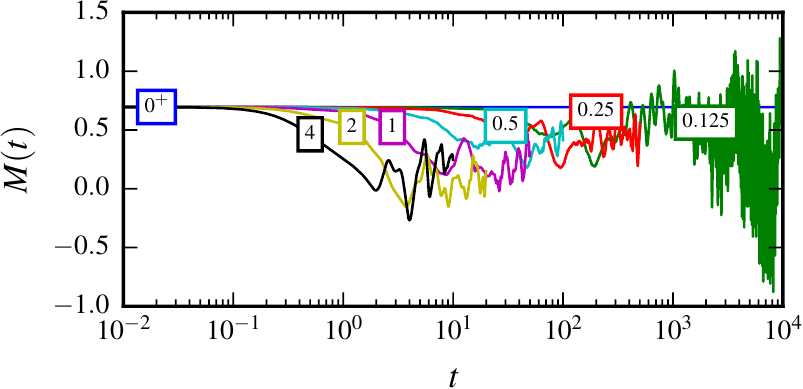}
    \caption{ (Color online) Violation of conservation of total magnetization
    $M$ by CPT.  The numbers indicate the value of $U_\mathrm{fin}$. Curves for
    $U_\mathrm{fin}\ge0.25$ are only partially plotted for better visibility.
 }
  \label{fig:total_density}
\end{figure}

The numerical results for the total energy, see Fig.\ \ref{fig:energy}, confirm
this expectation. Energy conservation is respected for $U_\text{fin}=0$, where
CPT is exact. With $U_{\rm fin}>0$ and increasing, however, a significant time
dependence of the total energy sets in earlier and earlier. For
$U_\text{fin}\gtrsim 1$ energy conservation is violated already for $t \lesssim
10$. Similar results are found for the total magnetization $M=\sum_{i}
(n_{i\uparrow}-n_{i\downarrow})$, cf.\ Fig.  \ref{fig:total_density}.  While
the magnetization should be constant for all times since neither hopping nor
interaction (cf.\ Eqs.\ \eqref{eq:cpt_hopping} and \eqref{eq:cpt_interaction})
involve spin-flip terms, we find such behavior only for short times. For longer
times oscillations arise and the conservation of total magnetization is
violated. For increasing $U_\mathrm{fin}$ the oscillations set in earlier
indicating again that the quality of CPT is best for values of $U_\mathrm{fin}$
close to zero.

We note that the total particle number $N = N_\uparrow + N_\downarrow$,
however, is conserved during the time evolution. This holds true for a
half-filled and homogeneously charged system and is due to the fact that CPT
preserves particle-hole symmetry. This can easily be understood as follows:
Each cluster Hamiltonian is particle-hole symmetric and since each cluster is
solved exactly within CPT the corresponding effective Hamiltonian $h^{I}(t)$
is also particle-hole symmetric. The CPT Hamiltonian is now given by Eq.
\eqref{eq:cpt_hamiltonian} which additionally includes the inter-cluster
hopping.  However, the inter-cluster hopping is clearly particle-hole symmetric
and so is the final CPT Hamiltonian.


\section{Summary and Outlook}
\label{sec:summary}
Concluding, we have shown that the nonequilibrium self-energy of an
interacting lattice-fermion model can uniquely be decomposed into a
superposition of noninteracting, isolated modes. This decomposition is a direct
analog to a well-established decomposition of equilibrium Green's functions,
called the \emph{Lehmann representation}. Our proof not only provides a direct
scheme to construct the Lehmann representation of the self-energy, and thus
allows for a deeper theoretical understanding of the self-energy complementary
to its diagrammatic definition, but also proves useful for practical
applications.

As a proof of concept we investigated the time evolution of local magnetic
moments in the fermionic Hubbard model after an interaction quench using
nonequilibrium cluster-perturbation theory. Our formalism allowed to avoid the
solution of an inhomogeneous Dyson equation on the Keldysh contour and we were
able to propagate the one-particle density matrix up to times
$t_\mathrm{max}=10^4$. 

On the physical side, quenches to weak $U_\mathrm{fin}$
turned out to be most interesting. In agreement with the predictions of general
perturbative considerations, \cite{moe:08, moe:09, moe:10,
  stk:13, eck:08.11} we found a separation of the dynamics into two
time scales. While the system qualitatively follows the constrained dynamics of
the non-interacting $U_\mathrm{fin}=0$ limit, the constraints are broken up for
large times due to the interaction and the system shows signs of relaxation.
However, memory of the initial state persists in the density matrix up to the
largest simulated times clearly indicating the absence of thermalization.

While the simple treatment of correlations by nonequilibrium CPT has shown to
be enough to cover the mentioned two-stage relaxation dynamics, it also leads
to a violation of the fundamental conservation laws of energy and total
magnetization. This could be fixed by additionally imposing a self-consistency
condition as it is done in nonequilibrium DMFT or in self-energy functional
theory.  Due to the significant, additional complexity of these approaches,
however, simulations would again be restricted to short time scales. A simpler,
more pragmatic approach might thus be preferable where, for example, local
continuity equations are enforced to ensure energy, total magnetization and
particle-number conservation. \cite{hof:13} Such a ``conserving
cluster-perturbation theory'' could allow for a complete dissipation of initial
perturbations and thus total loss of the memory of the initial state. Work
along these lines is in progress.

\begin{acknowledgments}
We thank Roman Rausch for providing an exact-diagonalization solver for the
Hubbard model, Felix Hofmann for a reference implementation of nonequilibrium
CPT, and Martin Eckstein and Karsten Balzer for helpful discussions.  This work
has been supported by the excellence cluster ``The Hamburg Centre for Ultrafast
Imaging - Structure, Dynamics and Control of Matter at the Atomic Scale'' and
by the Sonderforschungsbereich 925 (project B5) of the Deutsche
Forschungsgemeinschaft. Numerical calculations were performed on the PHYSnet
computer cluster at the University of Hamburg.
\end{acknowledgments}

\appendix
\section{Numerical construction of the effective Hamiltonian}
\label{ap:numerics}
\subsection{The $Q$-matrix and its time derivatives}
We assume that a small cluster is solved using exact diagonalization and that
all time derivatives $H^{(n)}(t) = \partial_t^n H(t)$ of the Hamiltonian are
known analytically. The numerical evaluation of Eq.\ \eqref{eq:qmat} for the
$Q$-matrix is straightforward within exact diagonalization. Its $n$-th
derivative can be obtained as follows. We have 
\begin{align}
  U^{(n)}(t,0) &= \partial_t^{n-1}(-i H(t) U(t,0))\\
 & =-i\sum_{k=0}^{n-1} \binom{n-1}{k}H^{(k)}(t)U^{(n-1-k)}(t,0)\nonumber
\end{align}
for the propagator $U(t,0)$. The $n$-th derivative $U^{(n)}(t,0)$ can then be
calculated iteratively as it only depends on $U^{(k)}(t,0)$ with $k<n$.  Using
further that
\begin{align}
  \partial_t^{n} \cc{i}(t) = \sum_{k=0}^{n} \binom{n}{k}
  U^{(k)}(t,0)\,\ccs{i}\,[U^{(n-k)}(t,0)]^\dagger 
  \: ,
\end{align}
one finds the $n$-th derivative $\cc{i}^{(n)}(t)$ of the annihilation operator
and thus of $Q^{(n)}(t)$, see Eq.\ \eqref{eq:qmat}. In the following we will
assume that $Q^{(n)}(t)$ is available to arbitrary order.

\subsection{Construction of the effective Hamiltonian at $t=0$}
We start by constructing $Q^\bot(0)$, i.e., a basis for the virtual sector. It
is easy to verify that
\begin{equation}
  P_{\alpha\alpha'}=\sum_{i} Q^*_{i\alpha}(0)Q_{i\alpha'}(0) \: ,
\end{equation}
defines a projector. Diagonalization of $P$ yields the eigenvalues $0$ and $1$.
Eigenvectors corresponding to $1$ are given by $Q(0)^\dagger$ itself,
eigenvectors corresponding to $0$ form the desired matrix
$[Q^\bot(0)]^\dagger$. Initially, the effective medium is in equilibrium and
thus explicitly given by Eq.\ \eqref{eq:heff_eq} at $t=0$.  However, since we
picked the completing basis vectors arbitrarily, we will have $h_{ss'} \neq 0$
for $s\neq s'$, i.e., generally $h$ will not be diagonal in the virtual sector.
Explicit diagonalization of $h$ in the virtual sector yields a unitary
transform $R$
\begin{equation}
  h_{ss'} = \sum_{r} R_{sr} d_{r} R^*_{rs'}.
\end{equation}
Replacing $Q^\bot(0)\rightarrow R Q^\bot(0)$, we get $h_{ss'}\rightarrow
\delta_{ss'}d_s$, i.e., we have found a completing basis so that $h$ is
diagonal in the virtual sector.

\subsection{The time derivatives $h^{(n)}(t)$}
Assume that $h(t), Q(t), Q^\bot(t)$ and $Q^{(n\ge 1)}(t)$ are known for an
arbitrary time $t$. This is at least the case for $t=0$ as we have seen so
far. We recall that we required $h(t)$ to be constant in the virtual sector
(cf.\ discussion below Eq.\ \eqref{eq:hermitian})
\begin{equation}
  h_{ss'}(t)=\delta_{ss'}h_{ss}(0)
  \quad\Rightarrow\quad h^{(n\ge 1)}_{ss'}(t) = 0,
\end{equation}
i.e., all time derivatives vanish in the virtual sector. Only the hybridization
elements and the physical sector yield non-trivial elements. They follow from
Eq.\ \eqref{eq:heff} as
\begin{align}
  \label{eq:hpv}
    h^{(n)}_{iy}(t)&=
    i\sum_{k=0}^{n} \binom{n}{k}
    \sum_\alpha [\partial_t^{k+1}(Q_{i\alpha}(t) e^{-i\epsilon_\alpha t})]
      [O^{(n-k)}(t)]^\dagger]_{\alpha y}.
\end{align}
$O^{(n)}(t)$ on the other hand only depends on $h^{(k)}(t)$, and $O^{(k)}(t)$,
for $k < n$, as readily follows from
\begin{align}
  O^{(n)}(t) &= -i\partial_t^{(n-1)} h(t) O(t)\\
             &= -i\sum_{k=0}^{n-1} \binom{n-1}{k} h^{(k)}(0) O^{(n-1-k)}(t).
             \nonumber
\end{align}
It is thus possible to iteratively calculate $O^{(n)}(t)$ and $h^{(n)}(t)$.

\subsection{Propagation of the $O$-matrix}
We assume that $O(t)$ and all derivatives of $h^{(n)}(t)$ are known at some
time $t$ and we want to propagate the $O$-matrix to $O(t+\Delta t)$.
Analytically this can be written as
\begin{equation}
  O(t+\Delta t) = \torder
    \left\{
      \exp{-i\int_t^{t+\Delta t} h(t')\dt'}
    \right\} O(t) \: .
\end{equation}
Using the Magnus expansion,\cite{bla:09} the propagator can be systematically
expanded in $\Delta t^n$ and $h^{(n)}(t)$. Assuming that $\Delta t$ lies within
the convergence radius of the Magnus expansion (this is generally expected to
be the case for sufficiently small $\Delta t$), we can reduce the propagation
error arbitrarily by increasing the order. In practice, an evaluation of the
Magnus expansion using commutator-free exponential time propagators
\cite{alv:11} (CFETs) allows for an efficient numerical propagation which takes
advantage of the sparse form of the effective Hamiltonian.

Having found $O(t+\Delta t)$, we get $h(t+\Delta t)$ from
\begin{align}
    h_{iy}(t+\Delta t)&= i\sum_\alpha Q^{(1)}_{i\alpha}(t+\Delta t) 
    e^{-i\epsilon_\alpha t}[O(t+\Delta t)^\dagger]_{\alpha y} \: ,
\end{align}
and can thus proceed by calculating $O^{(n)}(t+\Delta t)$ and
$h^{(n)}(t+\Delta t)$ completing the circle. We emphasize that the whole
procedure is numerically exact, i.e., the error is below machine precision, if
$\Delta t$ is chosen sufficiently small.


\begin{thebibliography}{44}
\expandafter\ifx\csname natexlab\endcsname\relax\def\natexlab#1{#1}\fi
\expandafter\ifx\csname bibnamefont\endcsname\relax
  \def\bibnamefont#1{#1}\fi
\expandafter\ifx\csname bibfnamefont\endcsname\relax
  \def\bibfnamefont#1{#1}\fi
\expandafter\ifx\csname citenamefont\endcsname\relax
  \def\citenamefont#1{#1}\fi
\expandafter\ifx\csname url\endcsname\relax
  \def\url#1{\texttt{#1}}\fi
\expandafter\ifx\csname urlprefix\endcsname\relax\def\urlprefix{URL }\fi
\providecommand{\bibinfo}[2]{#2}
\providecommand{\eprint}[2][]{\url{#2}}

\bibitem[{\citenamefont{Polkovnikov et~al.}(2011)\citenamefont{Polkovnikov,
  Sengupta, Silva, and Vengalattore}}]{pol:11}
\bibinfo{author}{\bibfnamefont{A.}~\bibnamefont{Polkovnikov}},
  \bibinfo{author}{\bibfnamefont{K.}~\bibnamefont{Sengupta}},
  \bibinfo{author}{\bibfnamefont{A.}~\bibnamefont{Silva}}, \bibnamefont{and}
  \bibinfo{author}{\bibfnamefont{M.}~\bibnamefont{Vengalattore}},
  \bibinfo{journal}{Rev. Mod. Phys.} \textbf{\bibinfo{volume}{83}},
  \bibinfo{pages}{863} (\bibinfo{year}{2011}).

\bibitem[{\citenamefont{Aoki et~al.}(2014)\citenamefont{Aoki, Tsuji, Eckstein,
  Kollar, Oka, and Werner}}]{aoki:14}
\bibinfo{author}{\bibfnamefont{H.}~\bibnamefont{Aoki}},
  \bibinfo{author}{\bibfnamefont{N.}~\bibnamefont{Tsuji}},
  \bibinfo{author}{\bibfnamefont{M.}~\bibnamefont{Eckstein}},
  \bibinfo{author}{\bibfnamefont{M.}~\bibnamefont{Kollar}},
  \bibinfo{author}{\bibfnamefont{T.}~\bibnamefont{Oka}}, \bibnamefont{and}
  \bibinfo{author}{\bibfnamefont{P.}~\bibnamefont{Werner}},
  \bibinfo{journal}{Rev. Mod. Phys.} \textbf{\bibinfo{volume}{86}},
  \bibinfo{pages}{779} (\bibinfo{year}{2014}).

\bibitem[{\citenamefont{Keldysh}(1964)}]{kel:64}
\bibinfo{author}{\bibfnamefont{L.~V.} \bibnamefont{Keldysh}},
  \bibinfo{journal}{J. Exptl. Theoret. Phys.} \textbf{\bibinfo{volume}{47}},
  \bibinfo{pages}{1515} (\bibinfo{year}{1964}).

\bibitem[{\citenamefont{Thygesen and Rubio}(2007)}]{thy:07}
\bibinfo{author}{\bibfnamefont{K.~S.} \bibnamefont{Thygesen}} \bibnamefont{and}
  \bibinfo{author}{\bibfnamefont{A.}~\bibnamefont{Rubio}},
  \bibinfo{journal}{The Journal of Chemical Physics}
  \textbf{\bibinfo{volume}{126}}, \bibinfo{pages}{091101}
  (\bibinfo{year}{2007}).

\bibitem[{\citenamefont{{Schmidt} and {Monien}}(2002)}]{sch:02}
\bibinfo{author}{\bibfnamefont{P.}~\bibnamefont{{Schmidt}}} \bibnamefont{and}
  \bibinfo{author}{\bibfnamefont{H.}~\bibnamefont{{Monien}}},
  \bibinfo{journal}{arXiv:cond-mat/0202046}  (\bibinfo{year}{2002}).

\bibitem[{\citenamefont{Freericks et~al.}(2006)\citenamefont{Freericks,
  Turkowski, and Zlati\ifmmode~\acute{c}\else \'{c}\fi{}}}]{fre:06}
\bibinfo{author}{\bibfnamefont{J.~K.} \bibnamefont{Freericks}},
  \bibinfo{author}{\bibfnamefont{V.~M.} \bibnamefont{Turkowski}},
  \bibnamefont{and}
  \bibinfo{author}{\bibfnamefont{V.}~\bibnamefont{Zlati\ifmmode~\acute{c}\else
  \'{c}\fi{}}}, \bibinfo{journal}{Phys. Rev. Lett.}
  \textbf{\bibinfo{volume}{97}}, \bibinfo{pages}{266408}
  (\bibinfo{year}{2006}).

\bibitem[{\citenamefont{Balzer and Potthoff}(2011)}]{bal:05.11}
\bibinfo{author}{\bibfnamefont{M.}~\bibnamefont{Balzer}} \bibnamefont{and}
  \bibinfo{author}{\bibfnamefont{M.}~\bibnamefont{Potthoff}},
  \bibinfo{journal}{Phys. Rev. B} \textbf{\bibinfo{volume}{83}},
  \bibinfo{pages}{195132} (\bibinfo{year}{2011}).

\bibitem[{\citenamefont{Jung et~al.}(2011)\citenamefont{Jung, Lieder, Brener,
  Hafermann, Baxevanis, Chudnovskiy, Rubtsov, Katsnelson, and
  Lichtenstein}}]{jung:11}
\bibinfo{author}{\bibfnamefont{C.}~\bibnamefont{Jung}},
  \bibinfo{author}{\bibfnamefont{A.}~\bibnamefont{Lieder}},
  \bibinfo{author}{\bibfnamefont{S.}~\bibnamefont{Brener}},
  \bibinfo{author}{\bibfnamefont{H.}~\bibnamefont{Hafermann}},
  \bibinfo{author}{\bibfnamefont{B.}~\bibnamefont{Baxevanis}},
  \bibinfo{author}{\bibfnamefont{A.}~\bibnamefont{Chudnovskiy}},
  \bibinfo{author}{\bibfnamefont{A.}~\bibnamefont{Rubtsov}},
  \bibinfo{author}{\bibfnamefont{M.}~\bibnamefont{Katsnelson}},
  \bibnamefont{and}
  \bibinfo{author}{\bibfnamefont{A.}~\bibnamefont{Lichtenstein}},
  \bibinfo{journal}{Ann. Phys.} \textbf{\bibinfo{volume}{524}},
  \bibinfo{pages}{49} (\bibinfo{year}{2011}).

\bibitem[{\citenamefont{Knap et~al.}(2011)\citenamefont{Knap, von~der Linden,
  and Arrigoni}}]{knap:11}
\bibinfo{author}{\bibfnamefont{M.}~\bibnamefont{Knap}},
  \bibinfo{author}{\bibfnamefont{W.}~\bibnamefont{von~der Linden}},
  \bibnamefont{and} \bibinfo{author}{\bibfnamefont{E.}~\bibnamefont{Arrigoni}},
  \bibinfo{journal}{Phys. Rev. B} \textbf{\bibinfo{volume}{84}},
  \bibinfo{pages}{115145} (\bibinfo{year}{2011}).

\bibitem[{\citenamefont{Hofmann et~al.}(2013)\citenamefont{Hofmann, Eckstein,
  Arrigoni, and Potthoff}}]{hof:13}
\bibinfo{author}{\bibfnamefont{F.}~\bibnamefont{Hofmann}},
  \bibinfo{author}{\bibfnamefont{M.}~\bibnamefont{Eckstein}},
  \bibinfo{author}{\bibfnamefont{E.}~\bibnamefont{Arrigoni}}, \bibnamefont{and}
  \bibinfo{author}{\bibfnamefont{M.}~\bibnamefont{Potthoff}},
  \bibinfo{journal}{Phys. Rev. B} \textbf{\bibinfo{volume}{88}},
  \bibinfo{pages}{165124} (\bibinfo{year}{2013}).

\bibitem[{\citenamefont{Joura et~al.}(2015)\citenamefont{Joura, Freericks, and
  Lichtenstein}}]{jou:15}
\bibinfo{author}{\bibfnamefont{A.~V.} \bibnamefont{Joura}},
  \bibinfo{author}{\bibfnamefont{J.~K.} \bibnamefont{Freericks}},
  \bibnamefont{and} \bibinfo{author}{\bibfnamefont{A.~I.}
  \bibnamefont{Lichtenstein}}, \bibinfo{journal}{Phys. Rev. B}
  \textbf{\bibinfo{volume}{91}}, \bibinfo{pages}{245153}
  (\bibinfo{year}{2015}).

\bibitem[{\citenamefont{Lipavsk\'y et~al.}(1986)\citenamefont{Lipavsk\'y,
  \ifmmode \check{S}\else \v{S}\fi{}pi\ifmmode~\check{c}\else \v{c}\fi{}ka, and
  Velick\'y}}]{lip:86}
\bibinfo{author}{\bibfnamefont{P.}~\bibnamefont{Lipavsk\'y}},
  \bibinfo{author}{\bibfnamefont{V.}~\bibnamefont{\ifmmode \check{S}\else
  \v{S}\fi{}pi\ifmmode~\check{c}\else \v{c}\fi{}ka}}, \bibnamefont{and}
  \bibinfo{author}{\bibfnamefont{B.}~\bibnamefont{Velick\'y}},
  \bibinfo{journal}{Phys. Rev. B} \textbf{\bibinfo{volume}{34}},
  \bibinfo{pages}{6933} (\bibinfo{year}{1986}).

\bibitem[{\citenamefont{Hermanns et~al.}(2014)\citenamefont{Hermanns,
  Schl\"unzen, and Bonitz}}]{bo:14}
\bibinfo{author}{\bibfnamefont{S.}~\bibnamefont{Hermanns}},
  \bibinfo{author}{\bibfnamefont{N.}~\bibnamefont{Schl\"unzen}},
  \bibnamefont{and} \bibinfo{author}{\bibfnamefont{M.}~\bibnamefont{Bonitz}},
  \bibinfo{journal}{Phys. Rev. B} \textbf{\bibinfo{volume}{90}},
  \bibinfo{pages}{125111} (\bibinfo{year}{2014}).

\bibitem[{\citenamefont{Weiss et~al.}(2008)\citenamefont{Weiss, Eckel,
  Thorwart, and Egger}}]{thor:08}
\bibinfo{author}{\bibfnamefont{S.}~\bibnamefont{Weiss}},
  \bibinfo{author}{\bibfnamefont{J.}~\bibnamefont{Eckel}},
  \bibinfo{author}{\bibfnamefont{M.}~\bibnamefont{Thorwart}}, \bibnamefont{and}
  \bibinfo{author}{\bibfnamefont{R.}~\bibnamefont{Egger}},
  \bibinfo{journal}{Phys. Rev. B} \textbf{\bibinfo{volume}{77}},
  \bibinfo{pages}{195316} (\bibinfo{year}{2008}).

\bibitem[{\citenamefont{Balzer and Eckstein}(2014)}]{eck:01.14}
\bibinfo{author}{\bibfnamefont{K.}~\bibnamefont{Balzer}} \bibnamefont{and}
  \bibinfo{author}{\bibfnamefont{M.}~\bibnamefont{Eckstein}},
  \bibinfo{journal}{Phys. Rev. B} \textbf{\bibinfo{volume}{89}},
  \bibinfo{pages}{035148} (\bibinfo{year}{2014}).

\bibitem[{\citenamefont{Fetter and Walecka}(2003)}]{wal:81}
\bibinfo{author}{\bibfnamefont{A.~L.} \bibnamefont{Fetter}} \bibnamefont{and}
  \bibinfo{author}{\bibfnamefont{J.~D.} \bibnamefont{Walecka}},
  \emph{\bibinfo{title}{Quantum Theory of Many-Particle Systems}}
  (\bibinfo{publisher}{Dover Publications}, \bibinfo{year}{2003}).

\bibitem[{\citenamefont{Gramsch et~al.}(2013)\citenamefont{Gramsch, Balzer,
  Eckstein, and Kollar}}]{cg:13}
\bibinfo{author}{\bibfnamefont{C.}~\bibnamefont{Gramsch}},
  \bibinfo{author}{\bibfnamefont{K.}~\bibnamefont{Balzer}},
  \bibinfo{author}{\bibfnamefont{M.}~\bibnamefont{Eckstein}}, \bibnamefont{and}
  \bibinfo{author}{\bibfnamefont{M.}~\bibnamefont{Kollar}},
  \bibinfo{journal}{Phys. Rev. B} \textbf{\bibinfo{volume}{88}},
  \bibinfo{pages}{235106} (\bibinfo{year}{2013}).

\bibitem[{\citenamefont{Stefanucci et~al.}(2014)\citenamefont{Stefanucci,
  Pavlyukh, Uimonen, and van Leeuwen}}]{ste:14}
\bibinfo{author}{\bibfnamefont{G.}~\bibnamefont{Stefanucci}},
  \bibinfo{author}{\bibfnamefont{Y.}~\bibnamefont{Pavlyukh}},
  \bibinfo{author}{\bibfnamefont{A.-M.} \bibnamefont{Uimonen}},
  \bibnamefont{and} \bibinfo{author}{\bibfnamefont{R.}~\bibnamefont{van
  Leeuwen}}, \bibinfo{journal}{Phys. Rev. B} \textbf{\bibinfo{volume}{90}},
  \bibinfo{pages}{115134} (\bibinfo{year}{2014}).

\bibitem[{\citenamefont{Gros and Valent{\'{\i}}}(1993)}]{gros:93}
\bibinfo{author}{\bibfnamefont{C.}~\bibnamefont{Gros}} \bibnamefont{and}
  \bibinfo{author}{\bibfnamefont{R.}~\bibnamefont{Valent{\'{\i}}}},
  \bibinfo{journal}{Phys. Rev. B} \textbf{\bibinfo{volume}{48}},
  \bibinfo{pages}{418} (\bibinfo{year}{1993}).

\bibitem[{\citenamefont{S{\'{e}}n{\'{e}}chal
  et~al.}(2000)\citenamefont{S{\'{e}}n{\'{e}}chal, Perez, and
  Pioro-Ladri{\`{e}}re}}]{sen:00}
\bibinfo{author}{\bibfnamefont{D.}~\bibnamefont{S{\'{e}}n{\'{e}}chal}},
  \bibinfo{author}{\bibfnamefont{D.}~\bibnamefont{Perez}}, \bibnamefont{and}
  \bibinfo{author}{\bibfnamefont{M.}~\bibnamefont{Pioro-Ladri{\`{e}}re}},
  \bibinfo{journal}{Phys. Rev. Lett.} \textbf{\bibinfo{volume}{84}},
  \bibinfo{pages}{522} (\bibinfo{year}{2000}).

\bibitem[{\citenamefont{S{\'{e}}n{\'{e}}chal
  et~al.}(2002)\citenamefont{S{\'{e}}n{\'{e}}chal, Perez, and
  Plouffe}}]{sen:02}
\bibinfo{author}{\bibfnamefont{D.}~\bibnamefont{S{\'{e}}n{\'{e}}chal}},
  \bibinfo{author}{\bibfnamefont{D.}~\bibnamefont{Perez}}, \bibnamefont{and}
  \bibinfo{author}{\bibfnamefont{D.}~\bibnamefont{Plouffe}},
  \bibinfo{journal}{Phys. Rev. B} \textbf{\bibinfo{volume}{66}},
  \bibinfo{pages}{075129} (\bibinfo{year}{2002}).

\bibitem[{\citenamefont{Jurgenowski and Potthoff}(2013)}]{jur:13}
\bibinfo{author}{\bibfnamefont{P.}~\bibnamefont{Jurgenowski}} \bibnamefont{and}
  \bibinfo{author}{\bibfnamefont{M.}~\bibnamefont{Potthoff}},
  \bibinfo{journal}{Phys. Rev. B} \textbf{\bibinfo{volume}{87}},
  \bibinfo{pages}{205118} (\bibinfo{year}{2013}).

\bibitem[{\citenamefont{Perfetti et~al.}(2006)\citenamefont{Perfetti, Loukakos,
  Lisowski, Bovensiepen, Berger, Biermann, Cornaglia, Georges, and
  Wolf}}]{perf:06}
\bibinfo{author}{\bibfnamefont{L.}~\bibnamefont{Perfetti}},
  \bibinfo{author}{\bibfnamefont{P.~A.} \bibnamefont{Loukakos}},
  \bibinfo{author}{\bibfnamefont{M.}~\bibnamefont{Lisowski}},
  \bibinfo{author}{\bibfnamefont{U.}~\bibnamefont{Bovensiepen}},
  \bibinfo{author}{\bibfnamefont{H.}~\bibnamefont{Berger}},
  \bibinfo{author}{\bibfnamefont{S.}~\bibnamefont{Biermann}},
  \bibinfo{author}{\bibfnamefont{P.~S.} \bibnamefont{Cornaglia}},
  \bibinfo{author}{\bibfnamefont{A.}~\bibnamefont{Georges}}, \bibnamefont{and}
  \bibinfo{author}{\bibfnamefont{M.}~\bibnamefont{Wolf}},
  \bibinfo{journal}{Phys. Rev. Lett.} \textbf{\bibinfo{volume}{97}},
  \bibinfo{pages}{067402} (\bibinfo{year}{2006}).

\bibitem[{\citenamefont{Bloch et~al.}(2008)\citenamefont{Bloch, Dalibard, and
  Zwerger}}]{bloch:08}
\bibinfo{author}{\bibfnamefont{I.}~\bibnamefont{Bloch}},
  \bibinfo{author}{\bibfnamefont{J.}~\bibnamefont{Dalibard}}, \bibnamefont{and}
  \bibinfo{author}{\bibfnamefont{W.}~\bibnamefont{Zwerger}},
  \bibinfo{journal}{Rev. Mod. Phys.} \textbf{\bibinfo{volume}{80}},
  \bibinfo{pages}{885} (\bibinfo{year}{2008}).

\bibitem[{\citenamefont{Rammer}(2007)}]{ram:07}
\bibinfo{author}{\bibfnamefont{J.}~\bibnamefont{Rammer}},
  \emph{\bibinfo{title}{Quantum Field Theory of Non-equilibrium States}}
  (\bibinfo{publisher}{Cambridge University Press},
  \bibinfo{address}{Cambridge, UK}, \bibinfo{year}{2007}).

\bibitem[{\citenamefont{van Leeuwen et~al.}(2006)\citenamefont{van Leeuwen,
  Dahlen, Stefanucci, Almbladh, and von Barth}}]{lee:06}
\bibinfo{author}{\bibfnamefont{R.}~\bibnamefont{van Leeuwen}},
  \bibinfo{author}{\bibfnamefont{N.~E.} \bibnamefont{Dahlen}},
  \bibinfo{author}{\bibfnamefont{G.}~\bibnamefont{Stefanucci}},
  \bibinfo{author}{\bibfnamefont{C.-O.} \bibnamefont{Almbladh}},
  \bibnamefont{and} \bibinfo{author}{\bibfnamefont{U.}~\bibnamefont{von
  Barth}}, \emph{\bibinfo{title}{Introduction to the Keldysh formalism}}, vol.
  \bibinfo{volume}{706} of \emph{\bibinfo{series}{Lecture Notes in Physics}}
  (\bibinfo{publisher}{Spinger}, \bibinfo{address}{Heidelberg, Germany},
  \bibinfo{year}{2006}).

\bibitem[{\citenamefont{Aichhorn et~al.}(2006)\citenamefont{Aichhorn, Arrigoni,
  Potthoff, and Hanke}}]{aich:06}
\bibinfo{author}{\bibfnamefont{M.}~\bibnamefont{Aichhorn}},
  \bibinfo{author}{\bibfnamefont{E.}~\bibnamefont{Arrigoni}},
  \bibinfo{author}{\bibfnamefont{M.}~\bibnamefont{Potthoff}}, \bibnamefont{and}
  \bibinfo{author}{\bibfnamefont{W.}~\bibnamefont{Hanke}},
  \bibinfo{journal}{Phys. Rev. B} \textbf{\bibinfo{volume}{74}},
  \bibinfo{pages}{235117} (\bibinfo{year}{2006}).

\bibitem[{\citenamefont{Metzner and Vollhardt}(1989)}]{voll:01.89}
\bibinfo{author}{\bibfnamefont{W.}~\bibnamefont{Metzner}} \bibnamefont{and}
  \bibinfo{author}{\bibfnamefont{D.}~\bibnamefont{Vollhardt}},
  \bibinfo{journal}{Phys. Rev. Lett.} \textbf{\bibinfo{volume}{62}},
  \bibinfo{pages}{324} (\bibinfo{year}{1989}).

\bibitem[{\citenamefont{Georges and Kotliar}(1992)}]{kot:03.92}
\bibinfo{author}{\bibfnamefont{A.}~\bibnamefont{Georges}} \bibnamefont{and}
  \bibinfo{author}{\bibfnamefont{G.}~\bibnamefont{Kotliar}},
  \bibinfo{journal}{Phys. Rev. B} \textbf{\bibinfo{volume}{45}},
  \bibinfo{pages}{6479} (\bibinfo{year}{1992}).

\bibitem[{\citenamefont{Potthoff}(2003)}]{po:03}
\bibinfo{author}{\bibfnamefont{M.}~\bibnamefont{Potthoff}},
  \bibinfo{journal}{Eur. Phys. J. B} \textbf{\bibinfo{volume}{32}},
  \bibinfo{pages}{429} (\bibinfo{year}{2003}).

\bibitem[{\citenamefont{Hofmann et~al.}(2015)\citenamefont{Hofmann, Eckstein,
  and Potthoff}}]{hof:15}
\bibinfo{author}{\bibfnamefont{F.}~\bibnamefont{Hofmann}},
  \bibinfo{author}{\bibfnamefont{M.}~\bibnamefont{Eckstein}}, \bibnamefont{and}
  \bibinfo{author}{\bibfnamefont{M.}~\bibnamefont{Potthoff}},
  \bibinfo{journal}{arXiv:1510.05866}  (\bibinfo{year}{2015}).

\bibitem[{\citenamefont{Kollath et~al.}(2007)\citenamefont{Kollath, L\"auchli,
  and Altman}}]{col:07}
\bibinfo{author}{\bibfnamefont{C.}~\bibnamefont{Kollath}},
  \bibinfo{author}{\bibfnamefont{A.~M.} \bibnamefont{L\"auchli}},
  \bibnamefont{and} \bibinfo{author}{\bibfnamefont{E.}~\bibnamefont{Altman}},
  \bibinfo{journal}{Phys. Rev. Lett.} \textbf{\bibinfo{volume}{98}},
  \bibinfo{pages}{180601} (\bibinfo{year}{2007}).

\bibitem[{\citenamefont{Trotzky et~al.}(2012)\citenamefont{Trotzky, Chen,
  Flesch, McCulloch, Schollw\"{o}ck, Eisert, and Bloch}}]{tro:12}
\bibinfo{author}{\bibfnamefont{S.}~\bibnamefont{Trotzky}},
  \bibinfo{author}{\bibfnamefont{Y.-A.} \bibnamefont{Chen}},
  \bibinfo{author}{\bibfnamefont{A.}~\bibnamefont{Flesch}},
  \bibinfo{author}{\bibfnamefont{I.~P.} \bibnamefont{McCulloch}},
  \bibinfo{author}{\bibfnamefont{U.}~\bibnamefont{Schollw\"{o}ck}},
  \bibinfo{author}{\bibfnamefont{J.}~\bibnamefont{Eisert}}, \bibnamefont{and}
  \bibinfo{author}{\bibfnamefont{I.}~\bibnamefont{Bloch}},
  \bibinfo{journal}{Nat. Phys.} \textbf{\bibinfo{volume}{8}},
  \bibinfo{pages}{325} (\bibinfo{year}{2012}).

\bibitem[{\citenamefont{Schneider et~al.}(2012)\citenamefont{Schneider,
  Hackerm\"uller, Ronzheimer, Will, Braun, Best, Bloch, Demler, Mandt, Rasch
  et~al.}}]{sch:12}
\bibinfo{author}{\bibfnamefont{U.}~\bibnamefont{Schneider}},
  \bibinfo{author}{\bibfnamefont{L.}~\bibnamefont{Hackerm\"uller}},
  \bibinfo{author}{\bibfnamefont{J.~P.} \bibnamefont{Ronzheimer}},
  \bibinfo{author}{\bibfnamefont{S.}~\bibnamefont{Will}},
  \bibinfo{author}{\bibfnamefont{S.}~\bibnamefont{Braun}},
  \bibinfo{author}{\bibfnamefont{T.}~\bibnamefont{Best}},
  \bibinfo{author}{\bibfnamefont{I.}~\bibnamefont{Bloch}},
  \bibinfo{author}{\bibfnamefont{E.}~\bibnamefont{Demler}},
  \bibinfo{author}{\bibfnamefont{S.}~\bibnamefont{Mandt}},
  \bibinfo{author}{\bibfnamefont{D.}~\bibnamefont{Rasch}},
  \bibnamefont{et~al.}, \bibinfo{journal}{Nature Physics}
  \textbf{\bibinfo{volume}{8}}, \bibinfo{pages}{213} (\bibinfo{year}{2012}).

\bibitem[{\citenamefont{Eckstein et~al.}(2010)\citenamefont{Eckstein, Kollar,
  and Werner}}]{eck:03.10}
\bibinfo{author}{\bibfnamefont{M.}~\bibnamefont{Eckstein}},
  \bibinfo{author}{\bibfnamefont{M.}~\bibnamefont{Kollar}}, \bibnamefont{and}
  \bibinfo{author}{\bibfnamefont{P.}~\bibnamefont{Werner}},
  \bibinfo{journal}{Phys. Rev. B} \textbf{\bibinfo{volume}{81}},
  \bibinfo{pages}{115131} (\bibinfo{year}{2010}).

\bibitem[{\citenamefont{Eckstein et~al.}(2009)\citenamefont{Eckstein, Kollar,
  and Werner}}]{eck:07.09}
\bibinfo{author}{\bibfnamefont{M.}~\bibnamefont{Eckstein}},
  \bibinfo{author}{\bibfnamefont{M.}~\bibnamefont{Kollar}}, \bibnamefont{and}
  \bibinfo{author}{\bibfnamefont{P.}~\bibnamefont{Werner}},
  \bibinfo{journal}{Phys. Rev. Lett.} \textbf{\bibinfo{volume}{103}},
  \bibinfo{pages}{056403} (\bibinfo{year}{2009}).

\bibitem[{\citenamefont{Moeckel and Kehrein}(2008)}]{moe:08}
\bibinfo{author}{\bibfnamefont{M.}~\bibnamefont{Moeckel}} \bibnamefont{and}
  \bibinfo{author}{\bibfnamefont{S.}~\bibnamefont{Kehrein}},
  \bibinfo{journal}{Phys. Rev. Lett.} \textbf{\bibinfo{volume}{100}},
  \bibinfo{pages}{175702} (\bibinfo{year}{2008}).

\bibitem[{\citenamefont{{Stark} and {Kollar}}(2013)}]{stk:13}
\bibinfo{author}{\bibfnamefont{M.}~\bibnamefont{{Stark}}} \bibnamefont{and}
  \bibinfo{author}{\bibfnamefont{M.}~\bibnamefont{{Kollar}}},
  \bibinfo{journal}{arXiv:1308.1610}  (\bibinfo{year}{2013}).

\bibitem[{\citenamefont{Rigol et~al.}(2008)\citenamefont{Rigol, Dunjko, and
  Olshanii}}]{rig:04.08}
\bibinfo{author}{\bibfnamefont{M.}~\bibnamefont{Rigol}},
  \bibinfo{author}{\bibfnamefont{V.}~\bibnamefont{Dunjko}}, \bibnamefont{and}
  \bibinfo{author}{\bibfnamefont{M.}~\bibnamefont{Olshanii}},
  \bibinfo{journal}{Nature} \textbf{\bibinfo{volume}{452}},
  \bibinfo{pages}{854} (\bibinfo{year}{2008}).

\bibitem[{\citenamefont{Moeckel and Kehrein}(2009)}]{moe:09}
\bibinfo{author}{\bibfnamefont{M.}~\bibnamefont{Moeckel}} \bibnamefont{and}
  \bibinfo{author}{\bibfnamefont{S.}~\bibnamefont{Kehrein}},
  \bibinfo{journal}{Annals of Physics} \textbf{\bibinfo{volume}{324}},
  \bibinfo{pages}{2146} (\bibinfo{year}{2009}).

\bibitem[{\citenamefont{Moeckel and Kehrein}(2010)}]{moe:10}
\bibinfo{author}{\bibfnamefont{M.}~\bibnamefont{Moeckel}} \bibnamefont{and}
  \bibinfo{author}{\bibfnamefont{S.}~\bibnamefont{Kehrein}},
  \bibinfo{journal}{New Journal of Physics} \textbf{\bibinfo{volume}{12}},
  \bibinfo{pages}{055016} (\bibinfo{year}{2010}).

\bibitem[{\citenamefont{Kollar et~al.}(2011)\citenamefont{Kollar, Wolf, and
  Eckstein}}]{eck:08.11}
\bibinfo{author}{\bibfnamefont{M.}~\bibnamefont{Kollar}},
  \bibinfo{author}{\bibfnamefont{F.~A.} \bibnamefont{Wolf}}, \bibnamefont{and}
  \bibinfo{author}{\bibfnamefont{M.}~\bibnamefont{Eckstein}},
  \bibinfo{journal}{Phys. Rev. B} \textbf{\bibinfo{volume}{84}},
  \bibinfo{pages}{054304} (\bibinfo{year}{2011}).

\bibitem[{\citenamefont{Blanes et~al.}(2009)\citenamefont{Blanes, Casas, Oteo,
  and Ros}}]{bla:09}
\bibinfo{author}{\bibfnamefont{S.}~\bibnamefont{Blanes}},
  \bibinfo{author}{\bibfnamefont{F.}~\bibnamefont{Casas}},
  \bibinfo{author}{\bibfnamefont{J.}~\bibnamefont{Oteo}}, \bibnamefont{and}
  \bibinfo{author}{\bibfnamefont{J.}~\bibnamefont{Ros}},
  \bibinfo{journal}{Physics Reports} \textbf{\bibinfo{volume}{470}},
  \bibinfo{pages}{151} (\bibinfo{year}{2009}).

\bibitem[{\citenamefont{Alvermann and Fehske}(2011)}]{alv:11}
\bibinfo{author}{\bibfnamefont{A.}~\bibnamefont{Alvermann}} \bibnamefont{and}
  \bibinfo{author}{\bibfnamefont{H.}~\bibnamefont{Fehske}},
  \bibinfo{journal}{Journal of Computational Physics}
  \textbf{\bibinfo{volume}{230}}, \bibinfo{pages}{5930} (\bibinfo{year}{2011}).

\end{thebibliography}
\end{document}